\documentclass{aa}
\usepackage{graphics}

\begin{document}

\thesaurus{03         
(11.03.2;  
11.05.2;  
11.06.1;  
11.09.2;  
11.19.3;  
11.19.4)} 
\title{
Dynamics of blue compact galaxies, \\
as revealed by their
H$\alpha$ velocity fields  : }

\subtitle{I. The data, velocity fields and rotation curves
\thanks{Based on observations collected
at the European Southern Observatory, La Silla, Chile}\fnmsep
\thanks{Table 6 is only available in electronic form at the CDS
 via anonymous ftp to cdsarc.u-strasbg.fr (130.79.128.5) or via
http://cdsweb.u-strasbg.fr/Abstract.html or from 
ftp://ftp.iap.fr/pub/from\_users/ostlin/Articles
}}

\author{G\"oran \"Ostlin \inst{1}\thanks{\emph{Present address:} Institut d'Astrophysique de Paris, F-75014 Paris, France}
 \and Philippe Amram \inst{2} 
\and Josefa Masegosa \inst{3} \and Nils Bergvall \inst{1} 
\and Jacques Boulesteix \inst{2}}
\offprints{G\"oran \"Ostlin, ostlin@iap.fr}
\institute{Astronomiska observatoriet, Box 515, S-75120 Uppsala, Sweden
\and
IGRAP, Observatoire de Marseille, 2 Place le Verrier, 
F-13248 Marseille Cedex 04, France
\and
Instituto de Astrofisica de Andalucia, CSIC, Apdo. 3004, E-18080 Granada, Spain}

\date{Received <date> / Accepted <date>}

\maketitle
\markboth{\"Ostlin et al.}{Dynamics of blue compact galaxies I. Observations }
\begin{abstract}
Observations of six luminous blue compact galaxies (BCGs) and two star forming companion galaxies were carried out with 
the CIGALE scanning Fabry-Perot	interferometer attached to the 
ESO 3.6m telescope on La Silla. The	observations were made in the H$\alpha$ 
emission line which is prominent in BCGs. A velocity sampling of
5 km/s and a pixel size of 0.9 arcseconds were used.  In this paper we present the observations and the data together with the velocity fields and the derived rotation curves. In addition we provide rough estimates of the total dynamical mass and of the ionised gas mass for each galaxy. All galaxies display rotation, but while the companion galaxies have regular velocity fields, those of the BCGs are complex and appear perturbed. This is the most extensive study to date of the optical velocity fields of BCGs. The interpretation of these results will be presented in a forthcoming paper (Paper II).  

\end{abstract}

\section{Introduction}
A Blue Compact Galaxy (BCG) is characterised by its blue optical colours, an \ion{H}{ii} region like emission line spectrum (therefore sometimes also referred to as an   \ion{H}{ii} galaxy) and compact appearance on photographic sky survey plates. BCGs have small to intermediate sizes (as measured e.g. by  $R_{25}$). Optical spectroscopy of BCGs in general reveal high star formation rates and low chemical abundances (Searle and Sargent \cite{searle:sargent}, Lequeux et al. \cite{lequeux:etal}, French \cite{french}, Kunth and Sargent \cite{kunth:sargent}, Mase\-gosa et al. \cite{masegosa}). Moreover most BCGs are rich in neutral hydrogen (Thuan and Martin \cite{thuan:martin}), a requisite for the intense star formation generally seen. However, the gas consumption time-scale is generally much shorter than the age of the universe, indicating that the high star formation rate must be transient. Such a galaxy is commonly referred to as a starburst galaxy. Thus BCGs are either genuinely young galaxies, or old galaxies that, for some reason, have resumed to form stars at a prodigious rate (Searle and Sargent  \cite{searle:sargent}, Searle et al. \cite{searle}). It is believed that BCGs undergo short (on the order of a few times 10 Myr) starbursts intervened by longer (on the order of a Gyr) passive periods.  A cyclic scenario has been proposed and may result from statistical effects (Searle et al. \cite{searle}, Gerola et al. \cite{gerola}). Another possibility is that supernova driven winds halt star formation by expelling the gas. Later, the lost gas might accrete back on the galaxy and create a new starburst (Dekel and Silk \cite{dekel}, Silk et al. \cite{silk}, and  Babul and Rees \cite{babul:rees}). Other ideas incorporate galaxy interactions as the triggering mechanism behind starburst activity (cf. e.g. Lacey et al. \cite{lacey}, Sanders et al. \cite{sanders}). Most BCGs are found outside galaxy clusters and in general  seems to be fairly isolated, although there are indications that HI-companions, sometimes without obvious optical counterparts, may be common (Taylor \cite{taylor}).

These different scenarios have been quite widely debated over the years, and while there is now ample evidence that most
BCGs contain old stars indicating that the present burst is  not the first one (for references see Paper II), there
is no consensus on the process(es) that trigger the bursts of star formation now evident. Most arguments have been based on photometry alone. On the other hand the dynamics of these systems are not well explored, still the creation of an energetic event like a sudden burst of star formation is likely to have dynamical causes and impacts.

To improve our understanding of the dynamics and the triggering mechanisms behind the starburst activity  we
have obtained H$\alpha$ velocity fields, using scanning Fabry-Perot (FP) interferometry, of a sample of BCGs.   With a FP it is possible to achieve a two dimensional velocity field with both high spatial and spectral (velocity) resolution. Thus we can get a much better view of the gas motions as compared to long slit spectroscopy. 
The velocity fields  can also be used to estimate the dynamical
masses of the galaxies, and from the H$\alpha$ intensities it is possible to estimate the mass of ionised gas.
Previous integral field studies of BCGs at high spectral resolution are rare. 
Thuan et al. (\cite{thuan:etal}) used Fabry-Perot interferometry to study two BCGs, both fainter than ours
(\object{VII Zw 403}, M$_B = -13.5$, and \object{I Zw 49}, M$_B = -17.3$).
While VII~Zw403 showed no well ordered large scale motion, I~Zw49 to some degree did.
Petrosian et al. (\cite{petrosian}) studied  \object{I Zw 18} ($M_B = -13.9$)  and found
indications of solid body rotation and asymmetric
line profiles, which they interpreted to be caused by gas
motions in the core of the galaxy.

%
%

\begin{table*}[tbp]
\caption[]{Journal of observations}
\label{table1}
\begin{flushleft}
\begin{tabular}[t]{lll}
\hline
\noalign{\medskip}
Observations & Telescope & ESO 3.6m\\
& Location & La Silla, Chile \\
& Equipment & CIGALE at Cassegrain focus \\   
& Date & August 30  to September 2, 1995\\   
& Seeing & $\sim 1.2 \arcsec$ \\  
Calibration 
& Neon line & $\lambda$ 6598.95 ~ \AA \\
Detector 
& Photon Counting Camera (IPCS) & Time resolution  1/50 s\\
Spatial Sampling 
& Pixel size & $0.91\arcsec$ \\
& Total Field & $230\arcsec \times 230\arcsec$ (256 $\times$ 256 pixels) \\
Temporal Sampling 
& Elementary scanning exposure time & 5 s per channel \\
\hline
Fabry-Perot interferometer no. 1 (FP1)
& Interference Order & 796 at  H$\alpha_0$ (6562.78 \AA)\\
& Free Spectral Range & 388 km/s \\
& Finesse at H$\alpha$ & 11 \\
& Number of Scanning Steps & 24\\
& Scanning Step & 0.36  \AA\ (16.15 km/s)\\
& Spectral resolution & $ R \ge 9000$ ~ (34 km/s)\\
\hline
Fabry-Perot interferometer no. 2 (FP2)
& Interference Order & 2604 at H$\alpha_0$ (6562.78 \AA) \\ 
& Free Spectral Range at H$\alpha_0$& 117 km/s \\
& Finesse at H$\alpha_0$ & 10 \\
& Number of Scanning Steps & 24\\
& Scanning Step & 0.105 \AA\ (4.80 km/s) \\
& Spectral resolution & $R \ge 26000$ ~ (12 km/s) \\
\noalign{\medskip}
\hline
\end{tabular}
\end{flushleft}
\end{table*} 

In this paper we will present the FP observations of  six luminous BCGs. These were selected to be bright in H$\alpha$ emission. Two of the galaxies have un-catalogued but confirmed star forming companion galaxies  and these were also observed. Most galaxies are from the sample by Berg\-vall and Olofsson (\cite{bergvall:olofsson}) and a newer extended version of it. In addition, one galaxy has been taken from the catalogue by Terlevich et al. (\cite{terlevich:etal}).   In this paper (Paper I) we will present the observations (Sect. 2), reductions (Sect. 3) and the results: the derived  H$\alpha$ images, velocity fields and  continuum images  (Sect. 4). In Sect. 5 we describe how the rotation curves (RCs) were constructed and provide rough mass estimates based on these. In Sect. 6 we give comments on the velocity fields and RCs of the individual target galaxies. In Sect. 7 we give a short summary of the results presented in this paper. Throughout this paper we will use a Hubble constant of $H_0 = 75$ km/s/Mpc.
In paper II (\"Ostlin et al. \cite{ostlin:cigale2}) we will discuss the interpretation of these results and their implications on the masses and dynamics of the galaxies and the triggering mechanism behind their  starbursts.

\section{Observations}
Eight galaxies were selected (see Sect. 1) for observations on the three dark nights
allocated at the ESO 3.6m telescope on La Silla,
from August 30 to September 2 1995.  
Observing conditions were photometric all nights. The seeing (measured at the telescope)
was slightly above one arcsecond.
The exposure times ranged between 24 minutes (1 minute per channel) and 160 minutes (almost 7 minutes per channel), plus calibration exposures.
The used instrument was CIGALE,
attached to the Cassegrain focus of the telescope.
CIGALE is basically composed of a focal reducer (bringing the original
f/8 focal ratio of the Cassegrain focus to f/3), a scanning
Fabry-Perot interferometer, an interference filter (to isolate the emission line and suppress the sky brightness)
and an IPCS detector (2-D photon-counting system).
The basic principles of this instrument were described in Amram et al.
(\cite{amram:1991}). The pixel size, projected on the sky, is 0.91$\arcsec$ with a resulting  field of view of $\sim$
4$\arcmin$.  The IPCS,
with a time resolution of 1/50 seconds and zero readout noise makes it
possible to scan the interferometer rapidly, avoiding problems with varying sky transparency, airmass 
and seeing  during long exposures; and thus has several advantages over a  CCD for this application. 
Narrow band interference filters were used to isolate the region around the redshifted  H$\alpha$ line. Different interference filters were used for observing galaxies  with different radial velocities, see Table \ref{table2}.
Only in two cases (\object{ES0 350-IG38} and \object{Tololo 0341-407})  the [\ion{N}{ii}]${\lambda 6548 \rm \AA}$ line might be partly transmitted. This line is however observed to be very weak as compared to H$\alpha$ in the two galaxies (Bergvall \& \"Ostlin \cite{bergvall:ostlin}, Terlevich et al.  
\cite{terlevich:etal}) so there is no risk that this affect our results.

In order to save observing time and to increase the
the spectral resolution,
with respect to the small velocity range of the observed galaxies, we
decided to use a high order Fabry-Perot interferometer 
(hereafter simply referred
as to FP2 for Fabry-Perot interferometer number 2, see Table \ref{table1})
giving at H$\alpha_0$ a mean finesse of 10, a free spectral range (FSR) of 117 km/s, and a scanning step of 4.8 km/s. The FSR is the separation between two consecutive interference orders and is therefore the maximum wavelength (velocity) range that can be observed before wavelength overlap occurs. 
The finesse is effectively the ratio between the FSR and the
width of the Airy function. The Airy function (or apparatus
function) is the instrumental line  profile. Thus by scanning the FSR in a 
number of steps that equals at least twice the finesse, adequate 
spectral sampling is obtained.
The spectral resolution, $R$,  given in Table 1 is calculated from the width of
the apparatus function, but for high signal to noise (S/N) the effective $R$
will be higher.

One galaxy was also observed using a lower order Fabry-Perot
interferometer (hereafter simply referred  to as
FP1, see Table \ref{table1}), which gave oss the possibility to check the consistency of our results. Calibrations were obtained by scanning a narrow neon line before and after the observations of each galaxy.
Table 1 lists some characteristics of the set-up used for
the observations. Table 2 gives the target names and  some parameters for the  observed
galaxies. 

%
%
\begin{table*}
\caption[]{List of observed galaxies. R.A. and  Decl. give the right ascension (h m s) and declination ($\degr ~ \arcmin ~ \arcsec$) of the targets for epoch 1950. $m_{\rm V}$ is the apparent magnitude in V.  $v_{\rm hel,obs}$ is the velocity used for selecting narrow band filter for the observations. "Filter" is the central wavelengths of the used interference filters as measured in laboratory and corrected to the observing conditions: a temperature of $\approx 5~\degr$C and a focal ratio of f/8. FWHM is the full with half maximum of the used narrow band filter. The transmission curve for the filters have a  quite squared shape, thus the full width at (practically) zero transmission is not much larger than the FWHM.  Unless noted otherwise, all observations were made with FP2.}
\label{table2}
\begin{flushleft}
\begin{tabular}{llllllllll}
\noalign{\smallskip}
\hline
\noalign{\smallskip}
Target Name & Other Name & R.A. & Decl. & $m_{\rm V}$ & $v_{\rm hel,obs}$ & Obs. date 
& Obs. time & Filter/{\small FWHM} & note/ \\
&& (1950) & (1950)	& & km/s & &	hours & \AA & ref. \\
\noalign{\smallskip}
\hline
\noalign{\smallskip}
\object{ESO 350-IG38} & \object{Haro 11}& 00 34 26 & -33 49 54 & 14.2  & 6175 & 30 Aug. 1995 & 2.27      &6691/22.5&1\\
\object{ESO 480-IG12} && 02 52 33	& -25 18 42 & 14.7   & 4710 & 31 Aug.  1995  & 2.43    &6670/20 &1\\
\object{ESO 338-IG04} & \object{Tololo 1924-416} & 19 24 29 & -41 40 24 & 13.6  & 2820 & 30 Aug. 1995  & 2.33 &6621/8.5 & 1,2\\
~~~~~~~~~"~~~ &      \object{SCHG 1924-416}               &        &           &        &  & 30 Aug. 1995  & 1.00 &&  1,3\\
\object{ESO 338-IG04B} && 19 24 03 & -41 45 00 &15.0  & 2925 & 31 Aug.  1995  & 2.40 &6630/9.5 & 9\\
\object{ESO 185-IG13} && 19 41 00  & -54 22 18 &15.1  & 5625 & ~1 Sept. 1995   & 2.66 &6691/22.5 & 4\\
\object{ESO 400-IG43} & \object{Fairall 1165}& 20 34 31 & -35 39 54 & 14.4  & 5830 & 30 Aug. 1995  & 2.33 &6691/22.5& 5\\
\object{ESO 400-IG43B} & & 20 34 41 & -35 40 12 & 15.7  & 5830 & 30 Aug. 1995  & 2.33 &6691/22.5& 5,6\\
\object{Tololo 0341-407}  & \object{SCHG 0341-407} & 03 41 03 & -40 45 29& 15.7   & 4500& ~1 Sept. 1995   & 0.40 &6653/20.5& 7,8 \\
\noalign{\smallskip} 
\hline
\end{tabular}\\
Notes/references:\\
 1. Position and velocity reference Bergvall and Olofsson (1986). \\ 
 2. Observation made with  FP2. \\
 3. Observation made with  FP1. \\
 4. From an extension of the Bergvall and Olofsson sample. Position and velocity reference from NED. \\ 
 5. \object{ESO 400-G43} and its companion \object{ESO 400-G43B} were observed 
simultaneously, since their separation is only 3 arcminutes. \\
 6. Companion to\object{ESO 400-G43}, position and velocity reference Bergvall and J\"ors\"ater (1988). \\
 7. Position and velocity reference Terlevich et al. (1991). \\
 8. Quoted magnitude is in the R band.  \\
 9. Companion to \object{ESO 338-IG04}, position and velocity reference Bergvall et al. (1998)\\
\end{flushleft}
\end{table*}

\section{Reductions}
Reduction of the data cubes were performed using the CIGALE/ADHOC software
(Boulesteix \cite{boulesteix}).  The data reduction procedure has been extensively
described in Amram et al. (\cite{amram:1995},\cite{amram:1996},\cite{amram:1997}) and references therein.
 The
accuracy of the zero point for the wavelength calibration is a fraction
of a channel width (${\rm <3  ~ km/s }$) over the whole field.
OH night sky lines passing through the filter were subtracted
by determining the emission in the field outside the galaxies (Laval et al. \cite{laval}).

Since we used a high order Fabry-Perot interferometer, FP2,
 the FSR of the interferometer
was of the same order of magnitude as the average H$\alpha$ line width
($\sim$ 100 km/s) of the target galaxies.  This means that in general we have no emission line free continuum.  Moreover, the wings of the emission lines
are shifted by a FSR (+1 FSR for the blueshifted wing and -1 FSR for
the redshifted wing) and superimposed on the central part of the line. However this does not
affect the velocity of the line, as long as the velocity range of the galaxy is lower than the FSR. When this is not the case, a velocity jump close to one FSR could easily be detected and corrected for using the continuity of the isovelocities.  The
only problem is that monochromatic (H$\alpha$) emission could be mistaken for continuum, which in effect would lower the measured line intensity and over-estimate the continuum level. To separate the continuum from the body and  wings of the
line, we used  a method based on  fitting 
theoretical line profiles  to the observed profiles.
The instrumental line width (apparatus function) of the interferometer 
has a FWHM of 10$\pm$2 km/s, where the range reflects the (known) variation of the FWHM over the field of view. 
Thus the observed widths of the profiles are dominated by the intrinsic H$\alpha$~ line widths in the  target galaxies. 
The convolution of the apparatus function with a theoretical H$\alpha$ profile yields something very close to a Gaussian.
The function used to fit to the observed profiles was a "folded" Gaussian: its wings were cut
(at $\pm$ one half FSR from the centre of the Gaussian),
shifted (by $\pm$1 FSR) and added to the main body of the profile. The reduced datacube provide the centre of the central wavelength,
and the sum of the continuum and monochromatic (H$\alpha$) emission.
Thus only two parameters are unknown: the width
and the amplitude of the Gaussian (i.e. the width and peak intensity of the H$\alpha$ line). These two parameters are
determined by a least-squares fit and thereby both the  continuum level and the H$\alpha$ intensity are determined for each pixel. 
To check the validity of this method, one galaxy (\object{ESO 338-IG04}) was observed
with two different interferometers having different
interference order: FP1 and FP2. 
Using FP1, the signal measured along the scanning
sequence is separated into
two parts: an almost constant level produced by the continuum light
passing through the  narrow band interference filter, and a varying part produced by the H$\alpha$ line emission.
The continuum level is the mean of all
channels which do not contain monochromatic (i.e. H$\alpha$) signal.
First of all, we checked that without any "data cooking",
the velocity fields have the same shape with
both FPs.  The agreement is even better if we degrade
the high resolution FP2 data to the spectral resolution of FP1, see also Sect. 6.3.
As a consequence of the lines having widths comparable to the FSR when using FP2, the continuum level and the true width of the lines are not known with accuracy; and therefore these quantities are not presented.

A rough flux calibration of the H$\alpha$ and continuum images of the galaxies could be made using the known instrumental sensitivity. For a narrow band filter having 100\%  transmission at the observed wavelength, one photo electron corresponds to a flux of $8.32 \times 10^{-17} \rm W m^{-2} s^{-1}$, with an uncertainty of $\pm 20 \%$. Correcting for the transmission of the narrow band filter and the atmospheric absorption we could then determine the total (line plus continuum) flux detected. As explained above, the separation of continuum and H$\alpha$ emission is not trivial.
However, in all cases the  H$\alpha$ emission dominate the detected flux. In effect the total H$\alpha$ fluxes are 
known with reasonable accuracy, while the absolute continuum level is uncertain. Comparing with available R-band photometry (e.g. Bergvall and Olofsson \cite{bergvall:olofsson}) we conclude that the contribution from the continuum to the detected flux is of the order of a few percent, which is slightly lower than the estimate form the continuum fit. In Table \ref{table5} the derived H$\alpha$ fluxes are presented. The quoted uncertainties represent the quadratic sum of the intrinsic uncertainty (20 \%) and the estimated uncertainty in determining the flux for each galaxy (10-30 \% due to uncertainty in the continuum level and the exact transmission of the filter at the wavelength of the redshifted H$\alpha$ line).  
The total derived H$\alpha$ fluxes can be used to estimate ${\cal M_{\rm ion}}$, the mass of the ionised H$\alpha$ emitting gas. For this we assume an electron density, $n_{\rm e}$, of 10 cm$^{-3}$, an electron temperature of 10\,000 K, a mean molecular weight, $\mu$, of 1.23 (to include the mass of helium) and use the H$\alpha$ effective recombination coefficient for case B as given by Osterbrock (\cite{osterbrock}).  The assumed density is in the lower range of observed densities of the \ion{H}{ii} gas in BCGs, which usually lie in the range $10 < n_{\rm e} < 300$ cm$^{-3}$ (see e.g. Bergvall \cite{bergvall}, and Masegosa et al. \cite{masegosa})  and thus our ${\cal M_{\rm ion}}$ estimate could be regarded as an upper limit, perhaps one order of magnitude too high.

\section{Fabry-Perot images and velocity fields}

In Fig. 1 to 14 we present the Fabry-Perot images of the observed galaxies.
In most cases a continuum image of each galaxy is given in the upper-left
panel, and a monochromatic (H$\alpha$) image in the mid-left panel.
The  contour levels  are usually the same
in the continuum and monochromatic
images, except that the continuum is the integrated intensity over the used narrow band filter while the monochromatic images present the total H$\alpha$ flux.
 The threshold of all the images was
chosen to be at 3$\sigma$ above the noise in the background. All images have linearly spaced contour levels.
The continuum images shown have been  smoothed with a Gaussian filter (FWHM of 3$\arcsec$). The  H$\alpha$~ 
images are the full resolution maps, but in drawing the contours the data was slightly smoothed with a median filter.
In general, in the right panel we present the
isovelocity contours of the ionised gas (thick solid lines)
superimposed on the H$\alpha$ image (thin solid lines).
The number of contour levels were here reduced
to make the plots more readable.
The heliocentric radial isovelocity lines have been drawn after a
smoothing the original profiles with a Gaussian filter (FWHM 1.8''~$\times$ 1.8''). 
The isovelocity contours are labelled in units of  km/s.
In each plot, the scale and orientation are given in the lower right corner.
In general the relative velocities are certain to within a fraction of the sampling step, but there might be systematic offsets of a few times the  FSR, since the  narrow band filters used are four to eight FSRs broad, giving a maximum systematic velocity error of a few times 100 km/s. However, for all galaxies there are independent velocity determinations available (see Table 2), which give us confidence in the determined systemic velocities (Table 3).

\subsection{Multicomponent decomposition} 

For some of the target galaxies we see obvious double line profiles (or profiles with very broad and flat peaks)   indicating that gas with two different velocities are present. When this was seen in more than a single isolated pixel but in a region of at least 3 by 3 pixels, and with sufficient S/N to rule out a noise peak, we attempted to decompose the velocity field into two components. This was done after subtracting the continuum and employing a small spectral smoothing with a Gaussian filter (FWHM of 15 km/s), in order to avoid 
artifacts due to noise. 
For this purpose a classical least-squares algorithm was used to fit two Gaussian profiles simultaneously to the full profile in the individual pixels. To reduce the number of free parameters in the fit, we imposed the constraint that the two Gaussians should have the same FWHM (as measured where they are well separated).    
The decomposition was performed individually on single pixels, but subject to a condition of continuity, i.e. in determining which of the fitted Gaussian components that belong to the same dynamical large scale component, the velocity of either component was not allowed to make sudden jumps from pixel to pixel. Thus the  velocity of one dynamical component in a certain pixel had to be consistent with its neighbouring pixels.

For several galaxies, we attempted a decomposition, but the results are only presented when we obtained a regular and continuous large scale secondary component. The failure to decompose the data cube into two components does not guarantee that secondary dynamical components are not present, only that we cannot detect them.
Moreover, we have no guarantee that the components are not 
separated by the value determined here plus the free spectral range
of the interferometer.  Nevertheless, the separation between the two
components are often so small than they should not be apparent if we
had used a lower resolution Fabry-Perot.
In those cases where we have extracted two dynamical components, individual H$\alpha$ profiles are superimposed on
the velocity field of the galaxy.
To make it more readable, one profile is usually drawn in a 2 by 2 pixels square box.
The abscissa of each small spectrum covers one FSR of
117~km/s, while the Y-amplitude is normalised to the peak value
of the monochromatic flux in the field of view. In some cases where we see double or broad lines, we also present the line profiles of the non-decomposed velocity field  with the Y-amplitude normalised to the maximum intensity in the box, which makes faint parts more visible.

For most galaxies, individual pixels or small regions can be found where the line profiles are double or asymmetric. This is an indication of small/intermediate scale gas motions, e.g. expanding bubbles or stellar winds. 

It cannot be excluded that several sources, overlaid on top of each other, have monochromatic emission within the narrow band filter (within a few times 100 km/s of the target galaxy). Then we could misinterpret a back\-ground/fore\-ground galaxy for a secondary dynamical component in the galaxy studied. This is however very unlikely in view of the appearance of the data. That the velocity structure within the target  galaxies (main component) could be a misinterpretation and rather be due to two separate galaxies a few 100 km/s apart is extremely unlikely since we then would expect to see double components in the profiles unless there was a perfect match in velocity (plus a multiple of the FSR), orientation  and  rotational velocity and direction.
Of course there might be more than two velocity components present, but we do not attempt a decomposition into more components, since third components do not display any meaningful structure over a scale larger than one or a few pixels. Thus third components (and second components in those cases where we do not decompose the velocity field) may reflect local small scale gas motions. A problem in attempting a multicomponent decomposition is that in principle any Gaussian can be decomposed into an infinite number of Gaussian components,  and thus the result would be unreliable and model dependent. 

%
%
%
\begin{table*}
\caption[]{
Parameters for  the rotation curves (RCs). For those galaxies where we have more than one dynamical component, or have drawn more than one RC, "Component" indicates which component the values refer to (see Sect. 6) and "RC" indicates the name of the corresponding RC.
The inclination ($Incl$) and position angle ($PA$) are 
the ones used in constructing (and iteratively determined from) the RCs. ~$S$ is the half sector (measured from the major axis) of the velocity field included in calculating the RCs. $PA_{\rm phot}$  and $Incl_{\rm phot}$ is the photometrically determined position angle and inclination respectively; the uncertainties are of the same order as for the kinematically determined properties. The absolute magnitudes and  ~$R_{25,B}$~ 
are based on the quoted radial velocities in Table \ref{table2} and  a Hubble parameter 
$H_0 = 75$ km/s/Mpc. $v_{\rm hel}$ is the determined systemic velocity in the heliocentric restframe which has a typical accuracy of 1 km/s. However $v_{\rm hel}$ may be systematically offset with ~$\pm$~ one FSR (117 km/s).  The $R_{25,B}$ values are corrected for inclination (using the photometric value) and given in arcseconds and  kpc. A colon after the value indicates that it is uncertain.
}
\begin{flushleft}
\begin{tabular}{llllllllllll} 
\noalign{\smallskip}
\hline
\noalign{\smallskip}
Target Name & Compo- & RC	& $v_{\rm hel}$ & $PA$  & $Incl$ & $S$  & $PA_{\rm phot}$ & $Incl_{\rm phot}$ & $R_{25,B}$ & $M_B$ & note \\
          &nent& Fig. & km/s  &	${\degr}$ & ${\degr}$  & ${\degr}$ & ${\degr}$ & ${\degr}$ & $\arcsec$/kpc & \\
\noalign{\smallskip}
\hline
\noalign{\smallskip}
\object{ESO 350-IG38}  &  Total    &  \ref{e350} & 6265 & 320 $\pm$ 10 & 40 $\pm$ 10 & 45 & 300 & 35    & 16.3/6.4 & -20.0 & 1\\
~~~~~~~"~~~~~ &  Main    &  \ref{e350_2} & 6264 & 320 $\pm$ 10 & 40 $\pm$ 10 & 45 &  &  & &  & 2\\
~~~~~~~"~~~~~ &  2:nd    &  \ref{e350_2} & 6250 & 140 $\pm$ 10 & 40 $\pm$ 25 & 85 & & &  &  & 3\\
\object{ESO 480-IG12}  & Main &  \ref{e480_1} & 4830 & 234 $\pm$ 10 & 52 $\pm$  5 & 60 & 200 & 64     & 19.5/5.9 & -19.1 & \\
\object{ESO 338-IG04}  & Main   &  \ref{e338_1}  & 2825 & ~60  $\pm$ 10 & 55 $\pm$ 10 & 45 & ~55  & 61  & 23.5/4.2 & -18.9 & 4\\
~~~~~~~"~~~~~ & Main   &  \ref{e338_3}  & 2816 & ~60  $\pm$ 10 & 55 $\pm$ 10 & 55 &   &   &  &  & 5\\
~~~~~~~"~~~~~ & Masked &  \ref{e338_2}a & 2815 & ~60  $\pm$ 10 & 55 $\pm$ 10 & 30 &     &     &  &  &6 \\
~~~~~~~"~~~~~ & 2:nd perp.&  \ref{e338_2}b & 2824 & 336  $\pm$ 5 & 62 $\pm$ 10 & 20 &  &     &  &  &7 \\
\object{ESO 338-IG04B} &        &  \ref{e338b} & 2900 & 230 $\pm$ 5  & 55 $\pm$ 5  & 40 & 225 & 58   & 18.2/3.3: & -17.5: &8 \\    
\object{ESO 185-IG13}  & Main & \ref{e185_1} & 5682 & ~25 $\pm$ 5 &  55 $\pm$ 5 & 40 & 130  & 42        & 12.4/4.5:& -18.7: &8 \\
 ~~~~~~~"~~~~ & 2:nd & \ref{e185_2} & 5706 & 225 $\pm$ 5 &  50 $\pm$ 10 & 40 &     &              &  &         & \\
\object{ESO 400-G43}   & &  \ref{e400} & 5813 & 225 $\pm$ 5  & 55 $\pm$ 10 & 55 & 220 & 40         & 14.3/5.3 & -19.8 & \\
\object{ESO 400-G43B}  & &  \ref{e400b} & 5886 & 305 $\pm$ 5  & 60 $\pm$ 5  & 45  & 310 & 73        & 13.0/4.9 & -18.3 & \\
\object{Tololo 0341-407}  & East &  \ref{tol_1} & 4535 & 295 $\pm$ 5 & 15 $\pm$ 5 & 60 &   &       & ~5.4/1.6: & -16.0:   &9 \\
 ~~~~~~~"~~~~ & West &  \ref{tol_2} & 4560 & 313 $\pm$ 5 & 50 $\pm$ 5 & 50 &   &       & ~8.8/2.5: & -17.0:   &9 \\
\noalign{\smallskip}
\hline
\end{tabular}\\
Notes: \\
1) Non-decomposed velocity field. \\
2) Primary component after decomposition of double line profiles in the centre. \\
3) Secondary component after decomposition of double line profiles in the centre. \\
4) Total velocity field from FP2 data. \\
5) Total velocity field from FP1 data. \\
6) FP2 data, parts of the velocity field have been masked, cf. Sect. 6.3.\\
7) Secondary perpendicular component, which has no obvious counterpart in broad band images, cf. Sect. 6.3. FP2 data. \\
8) $R_{25,B}$ and $M_B$ estimated from V-band data assuming $B-V=0.5$; this value is uncertain. \\
9) \object{Tololo 0341-407} has been decomposed into two components. Total absolute R-band magnitude: $M_R=-18.2$. The kinematical inclination and position angle
have been used also for calculating $R_{25,B}$. In addition $R_{25,B}$ has been estimated from an R-band  luminosity profile assuming $B-R=0.8$. Thus the values of $R_{25,B}$ for this object are uncertain. \\
\end{flushleft}
\end{table*}

\section{Rotation curves (RCs)}
In investigating the dynamics of the BCGs we prefer presenting a rotation curve (rather than just a position velocity cut) in order to indicate the real amplitude of the deprojected velocity field and to take into account data from a large fraction of the whole velocity field (and not only along a cut). However, we will see that the RCs are not always good tracers of the mass distribution in the galaxies.
The RC for each galaxy has been drawn by taking into account all velocity
points within $\pm S$ degrees (in the sky plane) from the kinematical major
axis (i.e. the $PA$).  This parameter was chosen to be as big as possible, still allowing a regular RC. The half sector, $S$,  used for each galaxy is indicated in the caption of the figure showing its RC and in Table 3.  
The RCs are given in Fig. 1 to 14 at the bottom of the pages. Table 6 gives the RCs in tabular form (this Table is only  published electronically)\footnote{Table 6 is only available in electronic form at the CDS
 via anonymous ftp to cdsarc.u-strasbg.fr (130.79.128.5) or via
http://cdsweb.u-strasbg.fr/Abstract.html, it is also available  at ftp://ftp.iap.fr/pub/from\_users/ostlin/Articles}. 
The ``cloud'' of small
points seen in each RC, are all the velocity points within ~$\pm 30 \degr$~ of the major axis in the sky plane. Although these points represent only a portion  
of the data (usually more data points are included in constructing the RC since normally ~$S > 30 \degr$), they give a fair idea of the amount, dispersion and quality of the data and the difference between the velocity on the major axis and the mean velocity. If the discrepancy is large, it means that circular motion is not likely.   
The error bars represent the $\pm 1 \sigma$~ dispersion in the velocity at each radius in the RC, and is a combination of the intrinsic dispersion and observational and reduction errors.
In general the intrinsic dispersion is larger than the observational errors. Thus the dispersion in the RC is in general caused by real irregularities in the velocity fields.   
The rotation velocity scale has not
been adjusted by the cosmological correction $(1+z)$.

The RCs assume axisymmetric objects with circular rotation and are sensitive to the choice of inclination ($Incl$), position angle ($PA$) and dynamical centre. The ~$PA$~ was determined from the orientation of the velocity field with a  typical accuracy of 5 to 10 degrees. The inclination was determined as to minimise the residuals and the dispersion in the RC. The centre coordinates and systemic velocity were chosen to give a RC with good agreement between the receding and approaching sides and to minimise the dispersion. The centre coordinates could be determined with a typical accuracy
of half a pixel and in general the displacement between the dynamical centre and the
broad band photometric centre (determined from the 1st statistical moment) is within one
pixel. 
The inclination is
in general the most uncertain parameter. Since many velocity fields look perturbed, a wide range
in ~$Incl$ may give comparably good fits overall. A typical accuracy for $Incl$ is 
$10\degr$. Secondary components have in general less well determined $Incl$.  The uncertainties in $Incl$  will enter in all mass estimates based on the RCs.

 The RC gives $V(R)$, the rotational velocity as a function  of radius, and in essence the rotational velocity is the velocity in the sky plane divided by $\sin(Incl)$. Thus the lower $Incl$, the more sensitive will the  derived velocities will be to uncertainties in $Incl$. The possible occurrence of warps or oval distortions and irregularities limit the validity of the RC, since its derivation is based on the assumption of a circular motions.

A rough estimate of the dynamical mass can be made using the simple model by Lequeux (1983) in which the mass within a radius $R$ is:

\begin{equation}
M(R)  = f \times R \times V^2(R)\times G^{-1}
\end{equation}

where $V(R)$ is the rotational velocity at $R$, ~$G$~ the gravitational constant and $f$ is a constant which has a value between 0.5 and 1.0. This formula is valid for any galaxy (in equilibrium) supported by rotation. For a disc with flat RC  $f=0.6$ while for a spherical distribution (e.g. a galaxy dominated by a dark halo) $f=1.0$. For a disc  with Keplerian decreasing RC  outside $R$, $f=0.5$ . Thus, according to this model, for any rotating galaxy $f$ should lie in the range: $f=0.5 {\rm ~to~} 1.0$, independent of the presence of a massive halo. Mass estimates based on this equation are given in Table 4, where we assumed $f=0.8$. The assumption  $f=0.8$ is not based on any physical reason, but was chosen simply to lie in the middle of the allowed interval. In addition to the intrinsic uncertainty in this  model, all the uncertainties above affects the accurateness of this estimate. In Paper II we will discuss these, and more refined, mass estimates of the observed galaxies. Of course, the mass estimate provided by Eq. (1), will only be a good approximation of the true dynamical mass if the galaxy is supported by rotation. If on the other hand the galaxy is mainly supported by random motions, the presented mass will be a severe underestimate.

The RC is based on the assumption of circular rotation. If this is not true, the RC will not give a good description of the dynamics and mass of  a galaxy. Nevertheless, even when the velocity field is perturbed and the derived RC looks weird (as for \object{ESO 338-IG04}, Fig. \ref{e338_1}), the RC gives some insights to the dynamics of a galaxy. Even the failure of constructing a symmetric and tight RC  is interesting, since it indicates that the system is complex or perturbed. 
In Table 3 we give some  parameters for the RCs. In some cases where we have extracted more than one component, we provide information for both. We also provide some photometric information, like $Incl_{phot}$ and $PA_{phot}$, the photometric inclination and position angle, respectively; and ~$R_{25,B}$, the radius at which the B surface brightness drops to 25 magnitudes per square arcsecond, corrected for inclination (cf. e.g. Bergvall et al. \cite{bergvall:etal}) and Galactic reddening. This information is not used in itself in the present investigation but provide complementary information. The  ~$R_{25,B}$~ is given in the RCs to give an idea of the size of the galaxy and the extent of the RC. When we did not have B-band data, we scaled V- or R-band data assuming crudely ~$B-V = 0.5$~ and $V-R=0.35$. The position angle and inclination derived from kinematics and photometry do not always agree which could be due to dynamical disturbances and instabilities in the systems or  internal absorption. Anyway, the only thing we used $PA_{phot}$ and $Incl_{phot}$ for in this investigation was to calculate ~$R_{25,B}$. Detailed surface photometry of some galaxies in this sample will be presented separately (Bergvall and \"Ostlin \cite{bergvall:ostlin}).

\begin{table}
\caption{Order of magnitude mass estimates. The first column gives the  target name. In those cases where the target seems to be composed of more than one dynamical component, column 2 indicates which component is considered: "Total" means that it is the total non-decomposed velocity field, "Main" means that it is the main component of the galaxy, "2:nd" that it is the secondary (normally weaker), and  "Comp" means that it is a companion galaxy.  Column 3 gives the  rotational velocity at the radius R, at which the mass is calculated.  Column 5 contains the mass estimate, ~${\cal M}$~, where we have assumed $f = 0.8$ in Eq. (1). See Sect. 5 for a discussion on the accuracy of these mass estimates. Sometimes there are more than one mass estimate for each galaxy (component); this is when the mass have been calculated at several radii, e.g. at the radius of maximum rotational velocity (indicated by $V_{max}$)  and at the last point in the RC (indicated by $R_{max}$), if these do not coincide. In the other cases, the maximum rotational velocity ($V_{max}$) occurs at the last point ($R_{max}$) in the RC and these are also the values of $V(R)$ and $R$ given in the Table below. The note "Both" indicates that the mass has been evaluated at the last point in the RC where there is data both from the receding and approaching sides. }
\label{table4}
\begin{flushleft} 
\begin{tabular}{llllll}
\hline
\noalign{\smallskip}
Target & Compo-     & $V(R)$ & $R$ & ${\cal M}$  & note \\
       & nent       & km/s            & kpc         & $10^6{\cal M_{\odot}}$ \\
\hline
\object{ESO 350-IG38}    & Total  &  95  &  0.2  & 340    &   $V_{max}$ \\
~~~~"~~~~~  	         & Total  &  30  &  5.4  & 900    &   $R_{max}$ \\
\object{ESO 350-IG38}    & Main   &  41  &  4.7  &1470   &  \\
\object{ESO 350-IG38}    & 2:nd   &  85  &  1.0  &1340   &  \\
\hline
\object{ESO 480-IG12}    & Main   &  120 &  2.8  & 7480   & Both\\
~~~~"~~~~~               & Main   &  140 &  4.5  & 16370  & $V_{max}$\\
~~~~"~~~~~               & Main   &  124 &  5.8  & 16550  & $R_{max}$\\
\hline
\object{ESO 338-IG04}    & Masked &  49  &  2.0  & 890    & $V_{max}$\\
\object{ESO 338-IG04}    & Masked &  35  &  2.5  & 570    & $R_{max}$\\
\object{ESO 338-IG04}    & 2:nd   &  19  &  0.8  & 54     & $V_{max}$ \\
 ~~~~"~~~~~              & ~"~~   &  12  &  1.9  & 51     & $R_{max}$ \\
\hline
\object{ESO 338-IG04B}   & Comp.  &  91  &  3.5  & 5380   & \\
\hline
\object{ESO 185-G13}     & Main   &  51  &  3.8  & 1830   & \\
\object{ESO 185-G13}     & 2:nd   &  26  &  2.7  & 340    & \\
\hline
\object{ESO 400-G43}     &        &  54  &  1.1  & 610    & $V_{max}$\\
~~~~"~~~~~               &        &  25  &  5.6  & 650    & Both \\
~~~~"~~~~~               &        &  12  &  10.2 & 270    & $R_{max}$\\
\hline
\object{ESO 400-G43B}    & Comp.  &  62  &  4.3  & 3070   & \\
\hline
\object{Tololo 0341-407} & East   &  24  &  1.6  & 170    & \\
\object{Tololo 0341-407} & West   &  27  &  1.9  & 260    & \\
\hline
\end {tabular}
\end{flushleft}
\end{table}

\begin{table}
\caption{Estimated H$\alpha$ fluxes  ($f_{\rm H\alpha}$), luminosities ($L_{\rm H\alpha}$) and  ionised gas masses(${\cal M_{\rm ion}}$). The H$\alpha$ luminosity was calculated using the quoted radial velocities in Table \ref{table2}. The mass was calculated assuming an electron density of $n_{\rm e} = 10$ cm$^{-3}$, an electron temperature $T_{\rm e} = 10\,000$K, and a mean molecular weight of $\mu = 1.23$. The H$\alpha$ recombination coefficient was taken from Osterbrock, Case B (\cite{osterbrock}).   
}
\label{table5}
\begin{flushleft}
\begin{tabular}{llll}
\hline
\noalign{\smallskip}
Target & $f_{{\rm H}\alpha}~~~~~~$  & $L_{{\rm H}\alpha}$ &  ~${\cal M_{\rm ion}}$  \\
       & 10$^{-17}$ W/m$^2$        & 10$^{33}$ W        & $10^6{\cal M_{\odot}}$    \\
\hline
\object{ESO 350-IG38}    & 350 $\pm$ 180 &  280 &  900  \\   
\object{ESO 480-IG12}    & 130 $\pm$  50 &   62 &   200 \\
\object{ESO 338-IG04}    & 290 $\pm$  60 &   49 &   160 \\
\object{ESO 338-IG04B}   &  24 $\pm$   5 &    4 &    14 \\
\object{ESO 185-G13}     & 120 $\pm$  35 &   81 &   260 \\
\object{ESO 400-G43}     & 240 $\pm$  70 &  170 &  550 \\
\object{ESO 400-G43B}    &  28 $\pm$   6 &   20 &    65 \\
\object{Tololo 0341-407 E}   &  45 $\pm$  13 &   19 &    61 \\
\object{Tololo 0341-407 W}   &  33 $\pm$   9 &   14 &    45 \\
\hline
\end {tabular}
\end{flushleft}
\end{table}

\section{Notes on the individual targets}
In this section we describe the images, velocity fields and rotation curves of the
individual targets. In paper II, the characteristics of the observed velocity fields and RCs will be analysed and interpreted.  In Table 3 we list the inclinations and position angles, derived from both photometry and kinematics. 
The kinematical centres have been compared with those from photometry, based on a first moment fit and the centre of outer isophotes. Unless noted otherwise, the agreement between the kinematic and photometric centres is within one pixel.

\subsection{\object{ESO 350-IG38}}
\begin{figure*}
\resizebox{14cm}{!}{\includegraphics{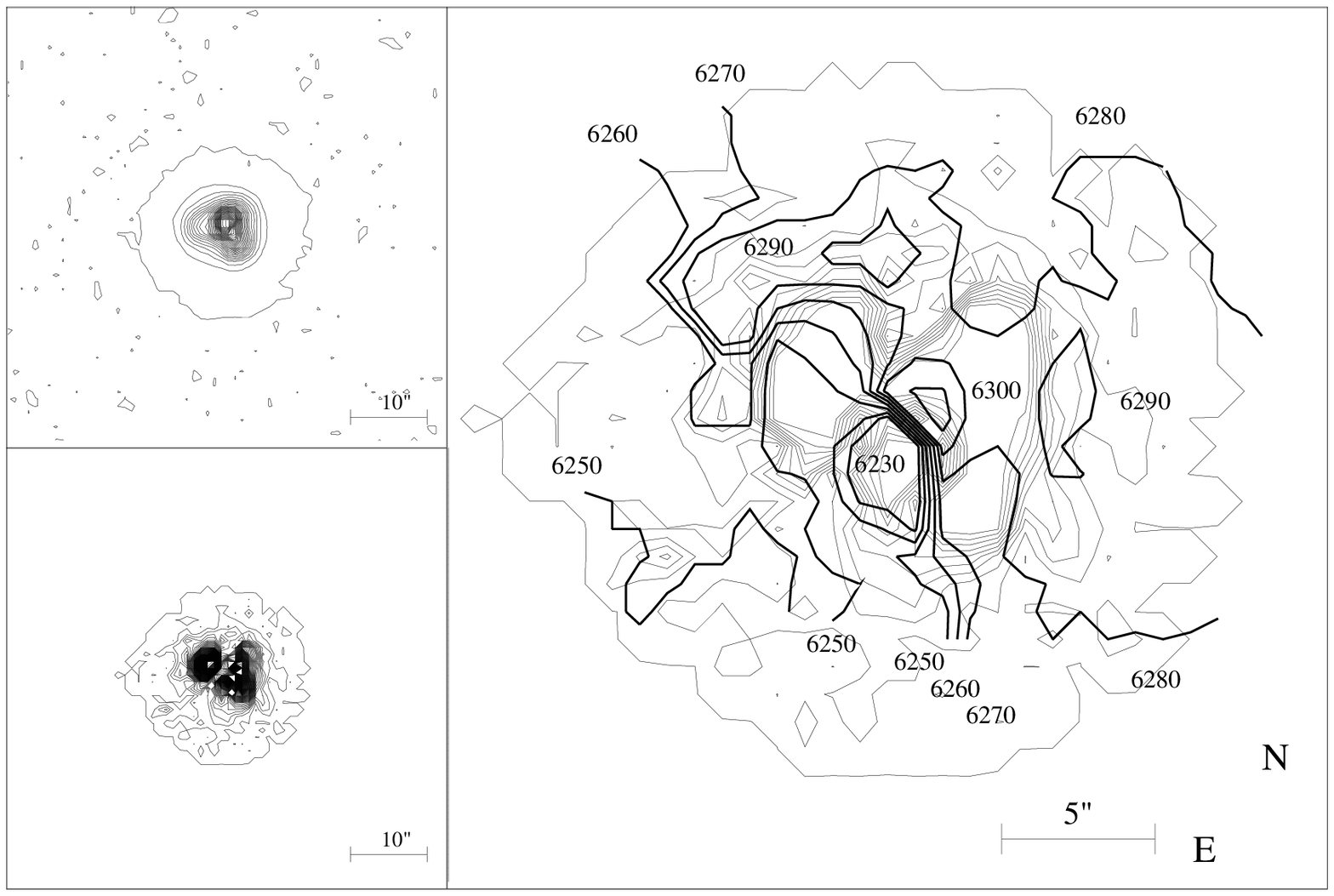}}
\resizebox{\hsize}{!}{\includegraphics{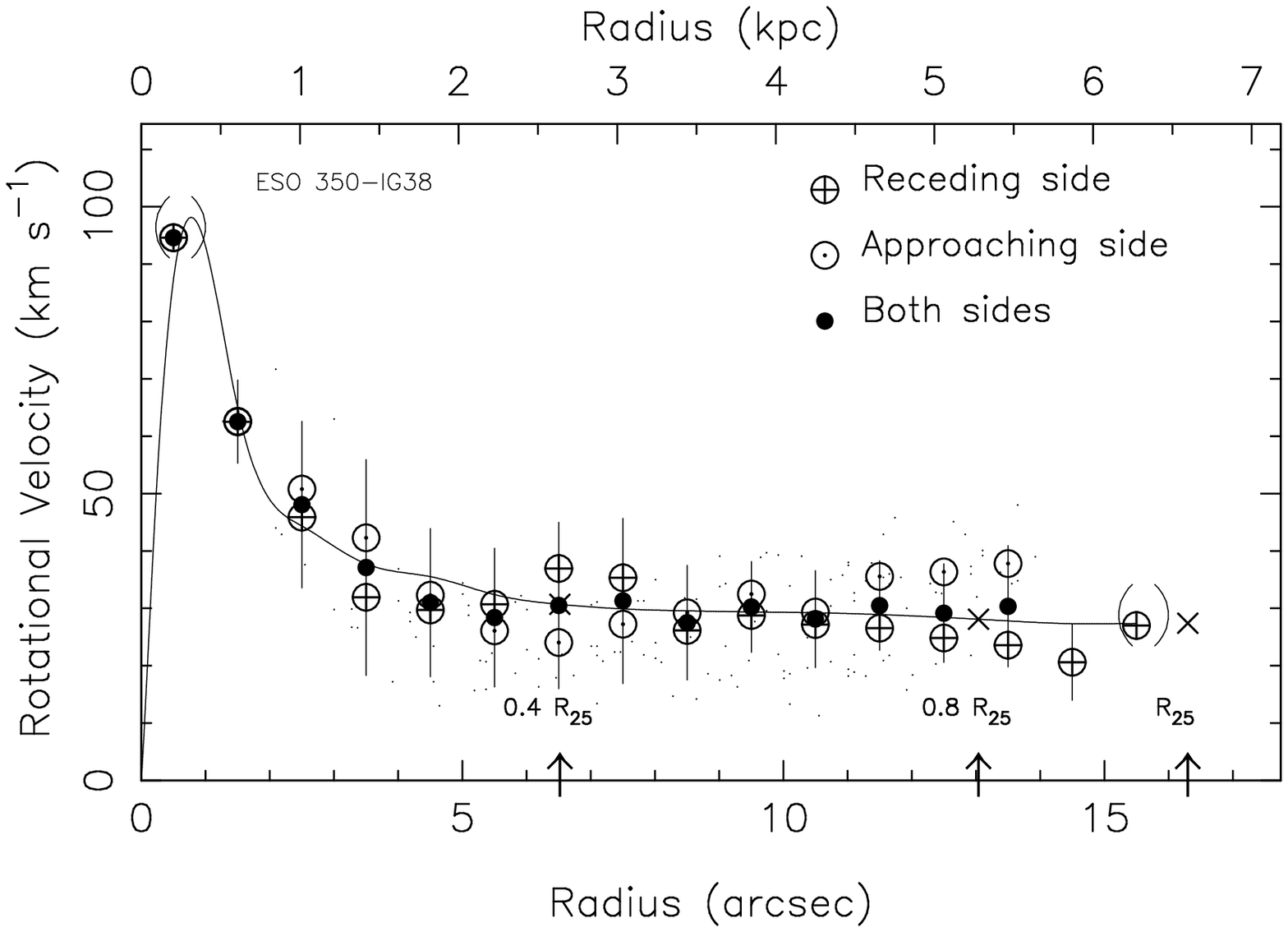}}
\caption{
\object{ESO 350-IG38}.
{\bf Upper left}: 
Continuum isointensity image.
{\bf Mid-left}: 
Monochromatic (H$\alpha$) isointensity map.  The (relative) step between two consecutive levels
is the same in both images.
{\bf Upper right}: 
Isovelocity contours of the ionised gas (thick solid lines)
superimposed on the H$\alpha$ image (thin solid lines).
The spatial scale in each image is shown in the lower right corner. North is up, east is to the left.  
{\bf Bottom}: Rotation curve (RC) derived from the velocity field and based on {\bf $ S = 45\degr$, $incl = 40\degr$~} and {\bf ~$PA = 320\degr$}. 
The receding side is indicated by $\bigoplus$ and the approaching side by  $\bigodot$. The  weighted average of the two sides is indicated 
by {\Large $\bullet$} with the lengths of the errorbars equal to the
$\pm 1 \sigma$ dispersion of the data points within the sector.
Velocity points which are based on only one or two data points are 
enclosed by brackets {\bf ()}. 
 The curve 
(solid line) is a smoothed splinefit to the average RC. The cloud of
 small dots represent those velocity points within $\pm 30 \degr$~from the major axis.
Arrows 
($\uparrow$) on the abscissa, and the crosses ($\times$) on the fitted RC,
mark the location of $0.4  R_{25}$, $0.8  R_{25}$, 
$1.0  R_{25}$, $1.2  R_{25}$. The scale on the upper abscissa is the 
radius in kpc (on the major axis), based on the radial velocity and H$_0 = 75$ km/s/Mpc.}
\label{e350}
\end{figure*}

\begin{figure*}
\resizebox{14cm}{!}{\includegraphics{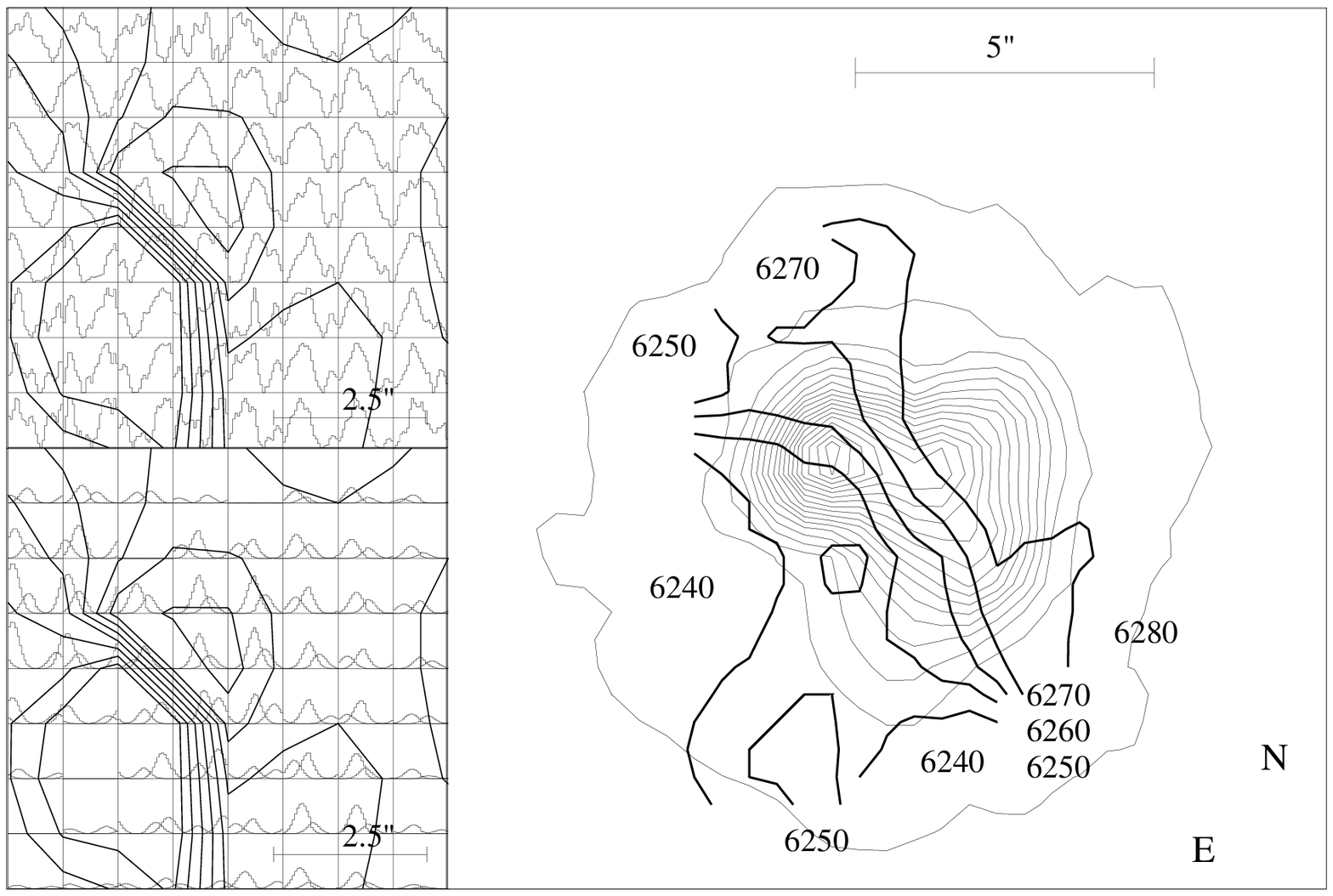}}
\resizebox{\hsize}{!}{\includegraphics{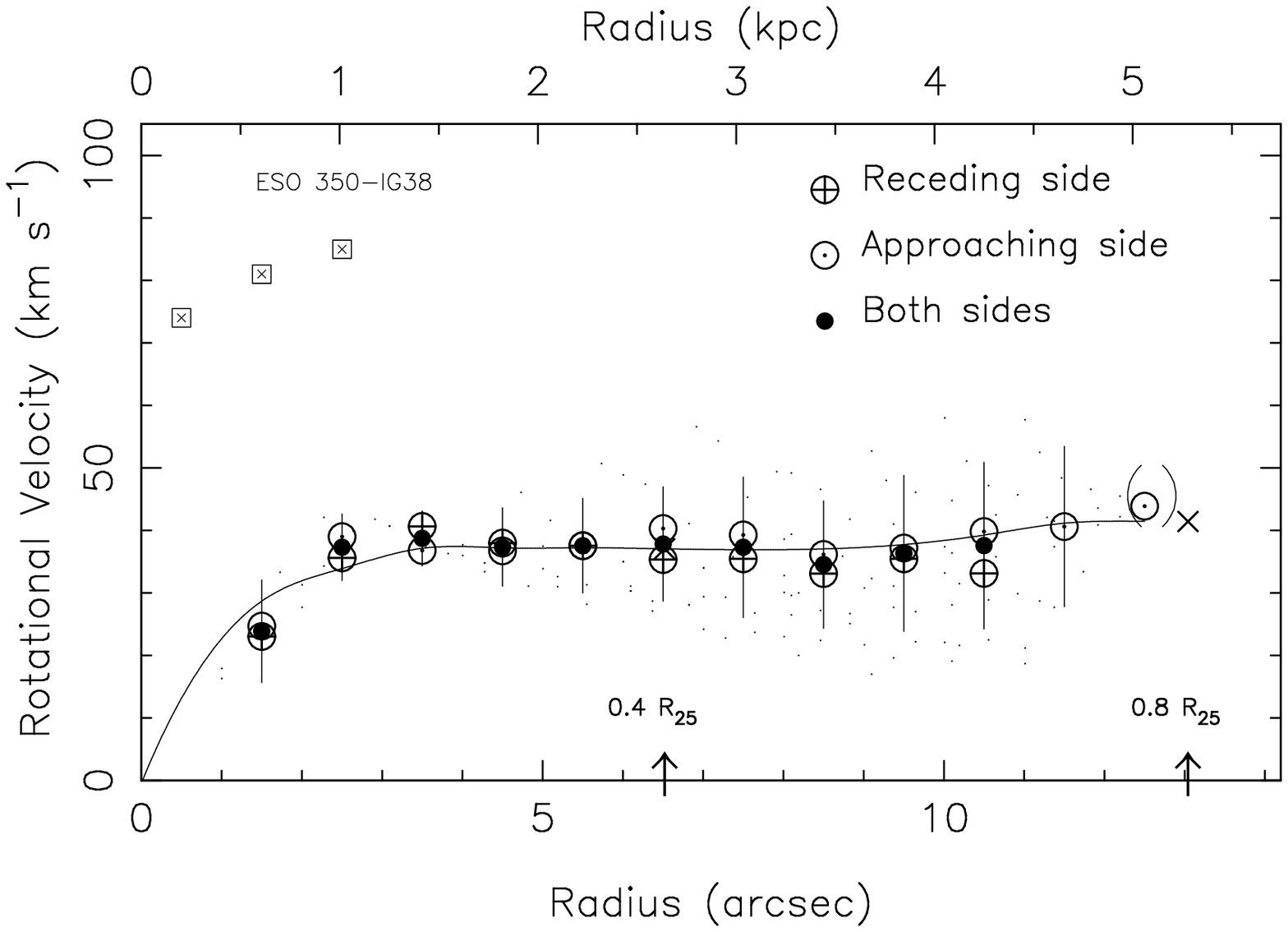}}
\caption{ \object{ESO 350-IG38} continued.
{\bf Upper left}:
 Line profiles in the centre of the non-decomposed velocity field plus isovelocity contours of the total, non-decomposed, velocity field. Each box corresponds to one  pixel. The Y-axis scale is normalised to the peak intensity in each pixel. 
{\bf Mid-left}:
 The line profiles in the centre of the two fitted components after decomposition of the velocity field. Each box corresponds to one pixel and the intensity scale is normalised to the pixel with the brightest H$\alpha$ flux.
{\bf Upper right}: Isovelocity contours of the first component of the ionised gas superimposed on the H$\alpha$ emission from the first component.
{\bf Bottom}:
  Rotation curve (RC)  based on the decomposed velocity field. The usual symbols
corresponds to the first component, and are based on $ S = 45\degr$, $incl = 40\degr$~ and ~$PA = 320\degr$. The secondary component, which is counter rotating, is shown as boxes with crosses in, and represent the weighted mean of both sides.  It is based on $ S = 85\degr$, $incl = 40\degr$~ and ~$PA = 140\degr$; N.B. that this RC is based on 9 pixels only. For further explanations see the caption of Fig. \ref{e350} and Sect. 5.
 }
\label{e350_2}
\end{figure*}

Our data (Fig. \ref{e350}, \ref{e350_2}) and broad band images  show that the morphology of 
this object is complex, apparently involving three nuclei/hot-spots. The outer 
H$\alpha$ and continuum isophotes appear fairly round. However, deep broad band 
images however reveal an  asymmetric morphology  at all radii. The kinematical centre of this galaxy lies between the north-west and south nuclei, approximately in the centre of the outer H$\alpha$ and continuum isophotes.
The isovelocity contours are very squeezed in the centre indicating a rapid increase in the rotational velocity. At larger radii (1 to 2 arcseconds)  a plateau is observed on each side followed by a velocity drop after which the RC stays flat at a rather low rotational velocity. Note that the steep central velocity gradient is not necessarily continuous, but could also be explained by H$\alpha$ emitting  blobs with different velocity tumbling  around each other.   
The deformations of the isovelocity contours (strongest in the north-east region) could indicate the presence of a spiral arm or a warp.
The emission lines of this galaxy are broad,
on the order of 200 km/s (Vader et al. \cite{vader}). Thus the full line profiles are much broader than our FSR, and the problems discussed in Sect. 3 are more serious 
for this galaxy.    

The derived RC  shows a rapid rise followed by
a Keple\-rian like decline shortly after, then levelling out at a constant velocity of 30 km/s for radii
larger than five arcseconds. Between approximately one and two kpc the RC falls slightly faster than the Keplerian prediction, but this is within the uncertainties. The receding and approaching sides are in reasonable agreement.

A close look a the line profiles in the centre revealed double lines
 which made us attempt a decomposition of the velocity field (see Fig
 \ref{e350_2}). This yielded a regular secondary component in the
 central part only.  After decomposing the velocity field the RC
 looks very different (see Fig. \ref{e350_2}). The central
 high velocity part in Fig.\ref{e350}  is replaced by a secondary
counter rotating component spinning 
at high velocity. To get good agreement between the receding and approaching
 sides of the secondary component, its systemic velocity was adjusted.
 Note that the RC of the secondary component is based on a total of nine pixels only.
The centre of the primary component nearly coincides with that of the
total non-decomposed velocity field. The primary component has
a nearly constant RC with a velocity of $\approx$  40 km/s. This  differs 
from the non-decomposed value of 30 km/s because it is based on fitted, i.e smoothed, data, and slightly different (1 pixel) centre coordinates. Moreover, since fitting  Gaussian profiles to the data requires higher S/N, the radial extent of this RC is somewhat smaller than for the non-decomposed velocity field.
A word of caution is necessary here: the velocity difference between 
the primary and secondary components is close to the FSR of the used
FP2, meaning that the lines almost overlap. In
effect the relative velocity of the secondary component with respect 
to the primary, is uncertain. Therefore we 
cannot exclude that the counter rotating  component is an artifact.
However there is a clear signature of asymmetric line profiles in the
centre and some sort of multicomponent gas is needed.
Observations with
a FP with higher FSR (e.g. FP1) should reveal if the second component
represents a counter rotating disc or not. At this stage we urge
the reader to view the RC of the second component in Fig. \ref{e350_2}
as one possible interpretation. If our interpretation is correct, the mass
of the secondary component is on the order of $10 ^9 \cal M_{\odot}$.

This galaxy has the greatest H$\alpha$ luminosity in the sample
 and the ionised gas mass may be as great as $10 ^9 \cal M_{\odot}$, 
 comparable to the dynamical mass estimate in Table 4.

\subsection{\object{ESO 480-IG12}}
This galaxy also has an overall irregular morphology. It is located in front of what appears to be a background cluster of galaxies. Three of the galaxies closest to \object{ESO 480-IG12} have measured redshifts that are much higher than that of \object{ESO 480-IG12}. Furthermore, none of
these objects show any H$\alpha$ emission at the redshift of \object{ESO 480-IG12}. Surprisingly though \object{ESO 480-IG12}
has faint warp-like extensions apparently aligned with a chain of background galaxies.

\begin{figure*}
\resizebox{15cm}{!}{\includegraphics{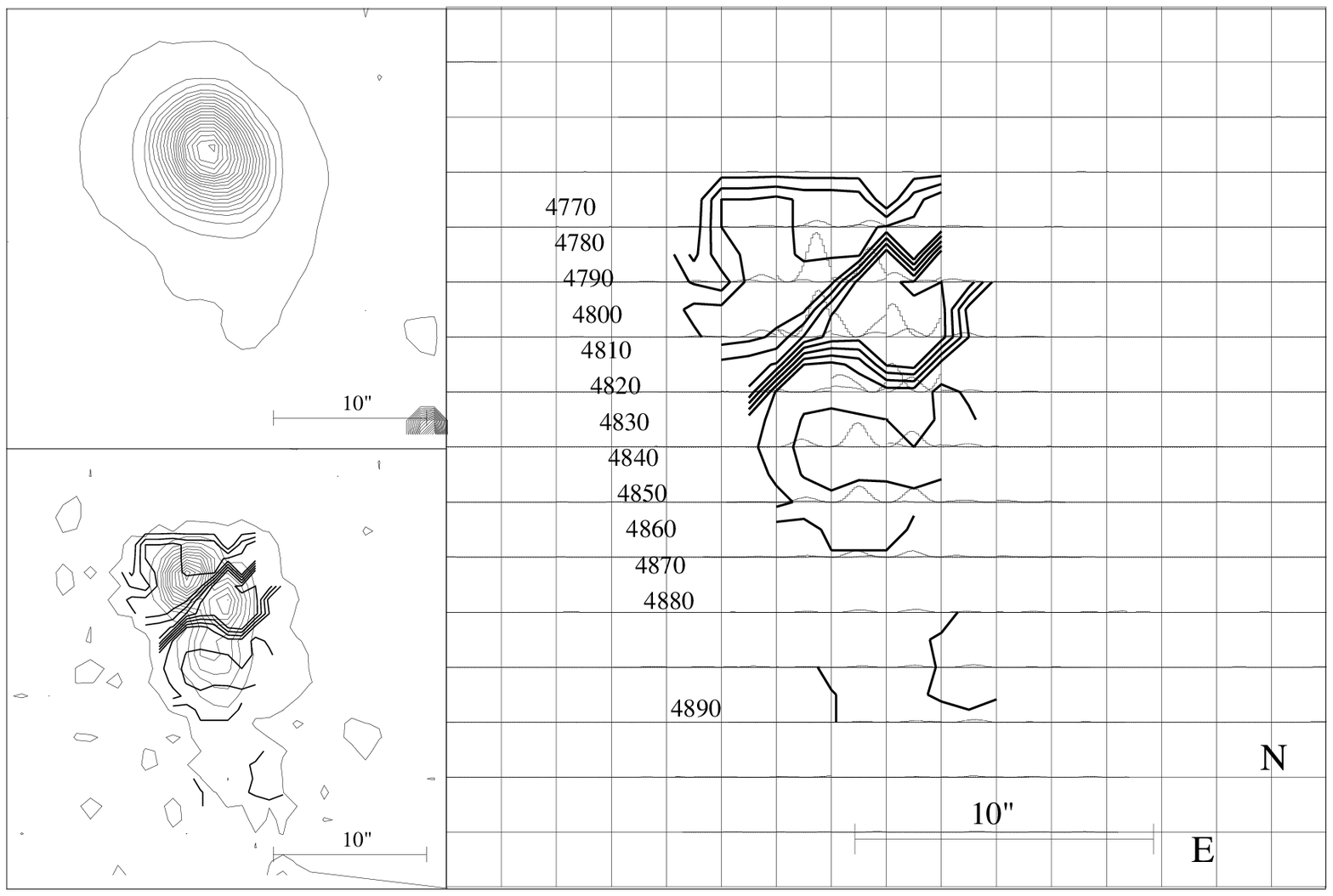}}
\resizebox{\hsize}{!}{\includegraphics{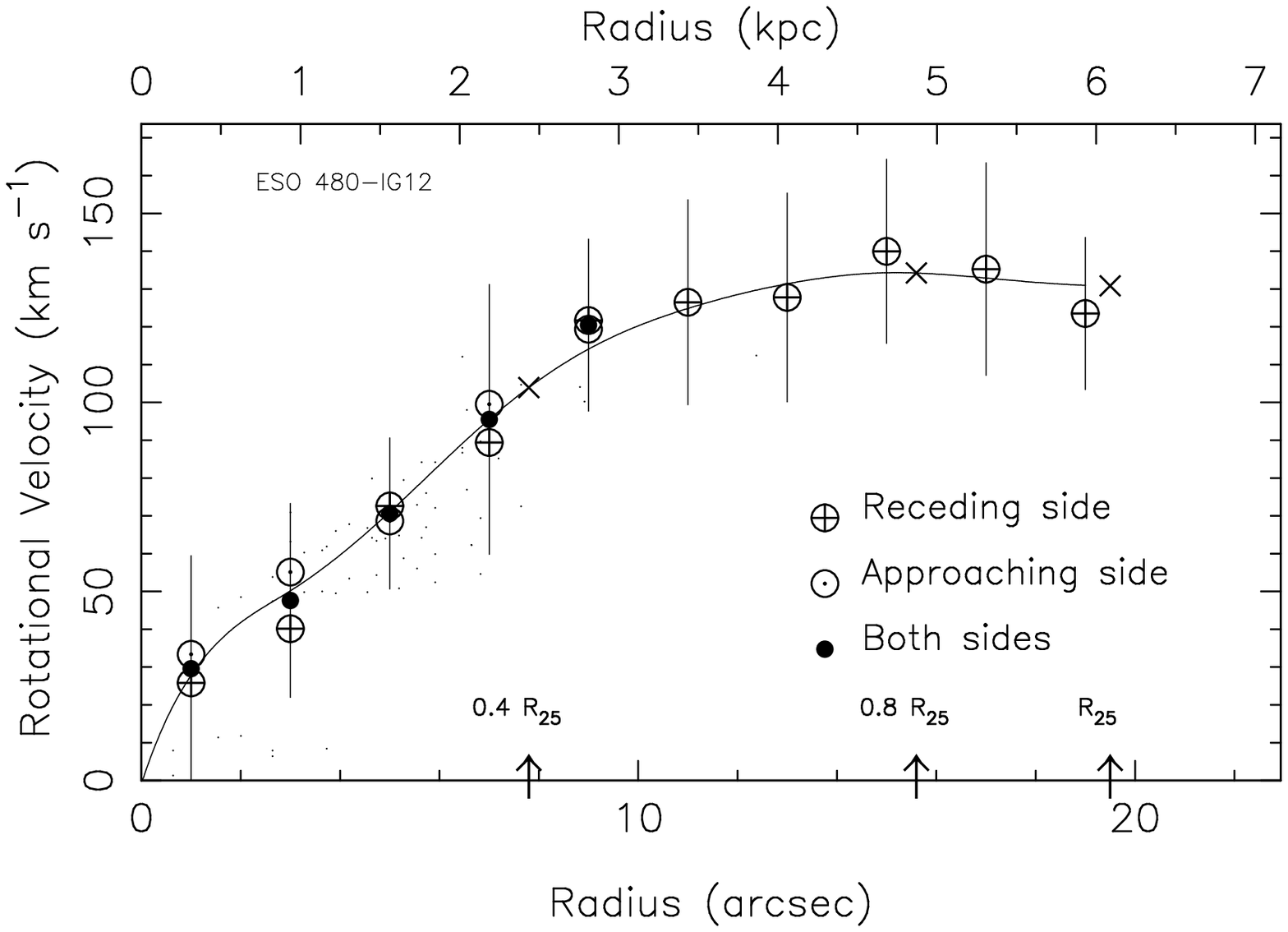}}
\caption{
\object{ESO 480-IG12}:
{\bf Upper left}: 
Continuum isointensity image.
{\bf Mid-left}: 
First monochromatic (H$\alpha$) component plotted
with the same step as in the continuum image. The isovelocity contours of the first 
component are overlaid (thick solid lines). 
{\bf Upper right}: 
Isovelocity contours of the first component of the ionised gas 
(thick lines) superimposed on the two Gaussian components of the two velocity fields 
(thin lines) plotted within squares of 2 pixels. The spatial scale in each image is shown in the lower right corner. North is up, east is to the left.
{\bf Bottom}:
 Rotation curve (RC)  based on $S = 60 \degr$, $incl = 52\degr$~ and ~$PA = 234\degr$. For further explanation of what is shown in the RC see the caption of Fig. \ref{e350} and Sect. 5.
}
\label{e480_1}
\end{figure*}

\begin{figure*}
\resizebox{\hsize}{!}{\includegraphics{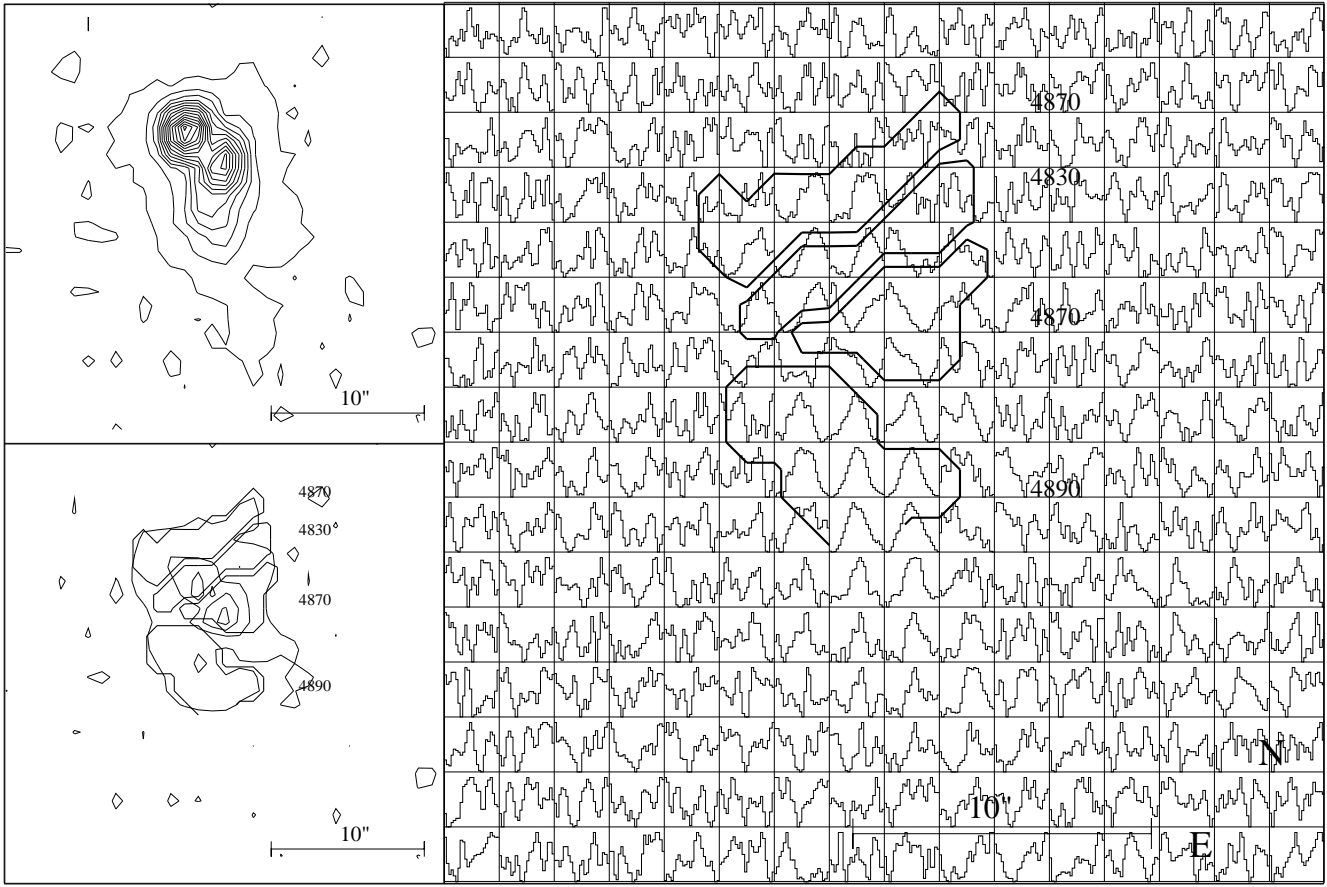}}
\caption{
\object{ESO 480-IG12} continued.
{\bf Upper left}: 
Monochromatic (H$\alpha$) isointensity image of the total (non-decomposed) velocity field.
{\bf Bottom left}: 
Second monochromatic component plotted
with the same step as in the continuum of the first component images (Fig. \ref{e480_1}). 
{\bf Right}: 
Isovelocity contours of the second component of the ionised gas 
(thick lines) superimposed on the non-decomposed line profiles  
(thin lines) plotted within squares of ~$2 \times 2$~ pixels. The intensity scale (Y-axis) in each 
box with line profiles is normalised to the peak intensity in that box. The spatial scale in each image is shown in the lower right corner. North is up, east is to the left.
}
\label{e480_2}
\end{figure*}

The H$\alpha$ emission
is concentrated to two, or perhaps three,  central \ion{H}{ii} regions. 
In between the \ion{H}{ii} regions the isovelocity contours are squeezed.
Double and very broad components are observed in
the emission line profiles, especially west of the centre (see Fig. \ref{e480_1} and \ref{e480_2}). Two components have been extracted with quite
different intensity levels.  If we do not decompose the 
profiles and simply compute the velocity from the 
total H$\alpha$ line, the first component clearly dominates in the brightest part,
but in fainter regions west of the centre, unreasonable features appear like the confusing situation with three different symmetry axes. Anyway, this would not affect the RC of the primary component since the regions with double lines are close to the minor axis. For the primary component, the velocity field shows a strong 
gradient along the major axis; but with a plateau in the north-west region. This coincides with the western \ion{H}{ii} region and roughly with the region where the double features in the lines are most pronounced, and thus close to where the second component has its maximum intensity.
The kinematical centre is well defined along the major axis and coincides roughly with 
the continuum peak intensity. Due to the plateau, the kinematical centre
is less well determined along the minor axis, however different choices of 
 centre along this axis gives consistent RCs. In essence the RC does not sensitively depend on the centre coordinates or the decomposition model.

The RC derived for the main component shows solid body rotation
out to a radius of $\approx 3$ kpc, after which it levels out, although it is here based on the receding side only. The RC has a rather large dispersion indicating that the physical situation might be  more complex than the assumed 
disk anatomy. For the second component, we could not obtain any sensible RC. However, the duality of the velocity peaks in the western part of
the galaxy are very significant, and some sort of multicomponent gas is needed
to explain the data. The ionised gas mass is of the order $10^8 \cal  M_{\odot}$, thus significantly smaller than the dynamical mass estimate which is of the order $  10^{10} \cal M_{\odot}$.

\subsection{\object{ESO 338-IG04}}
\begin{figure*}
\resizebox{15cm}{!}{\includegraphics{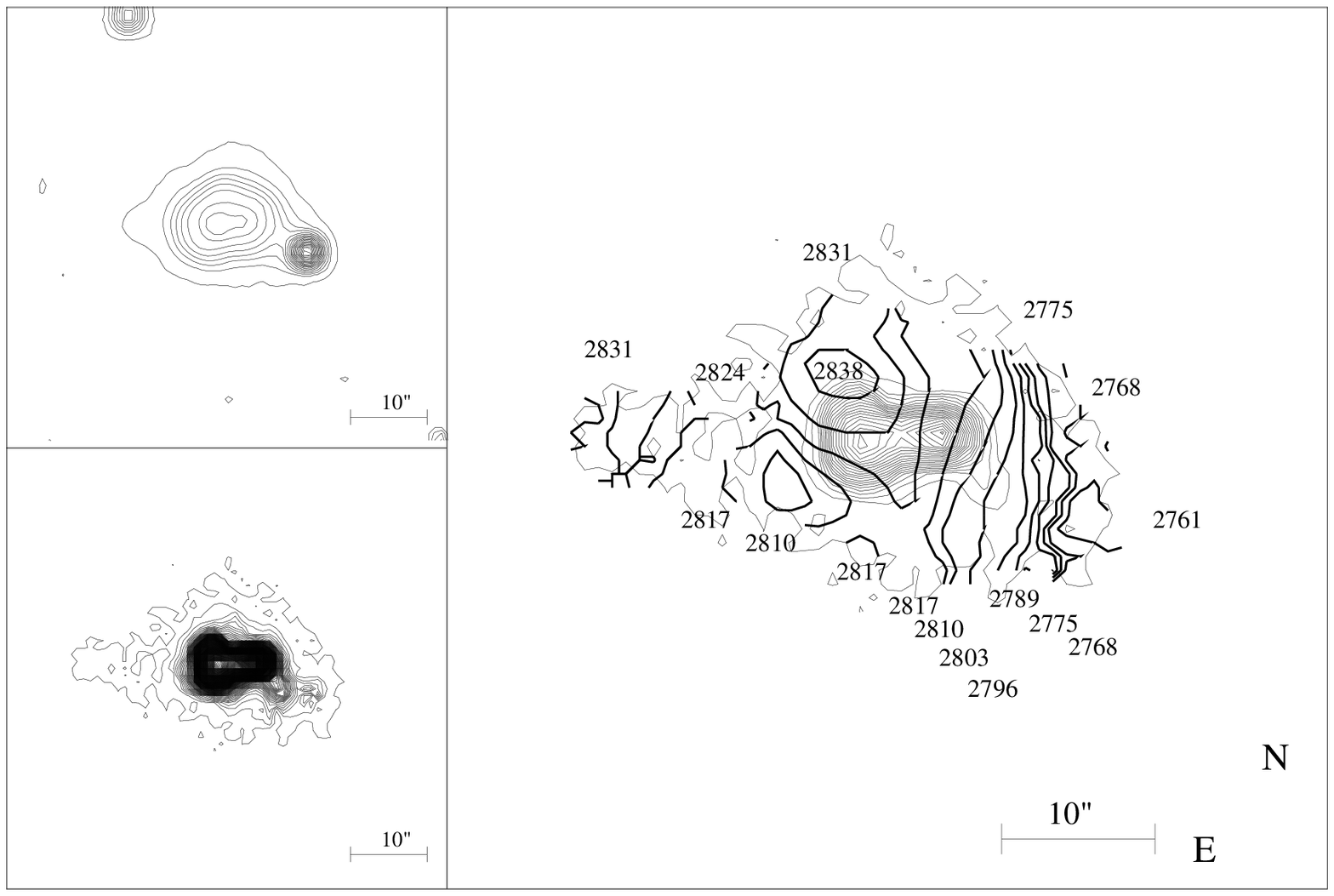}}
\resizebox{\hsize}{!}{\includegraphics{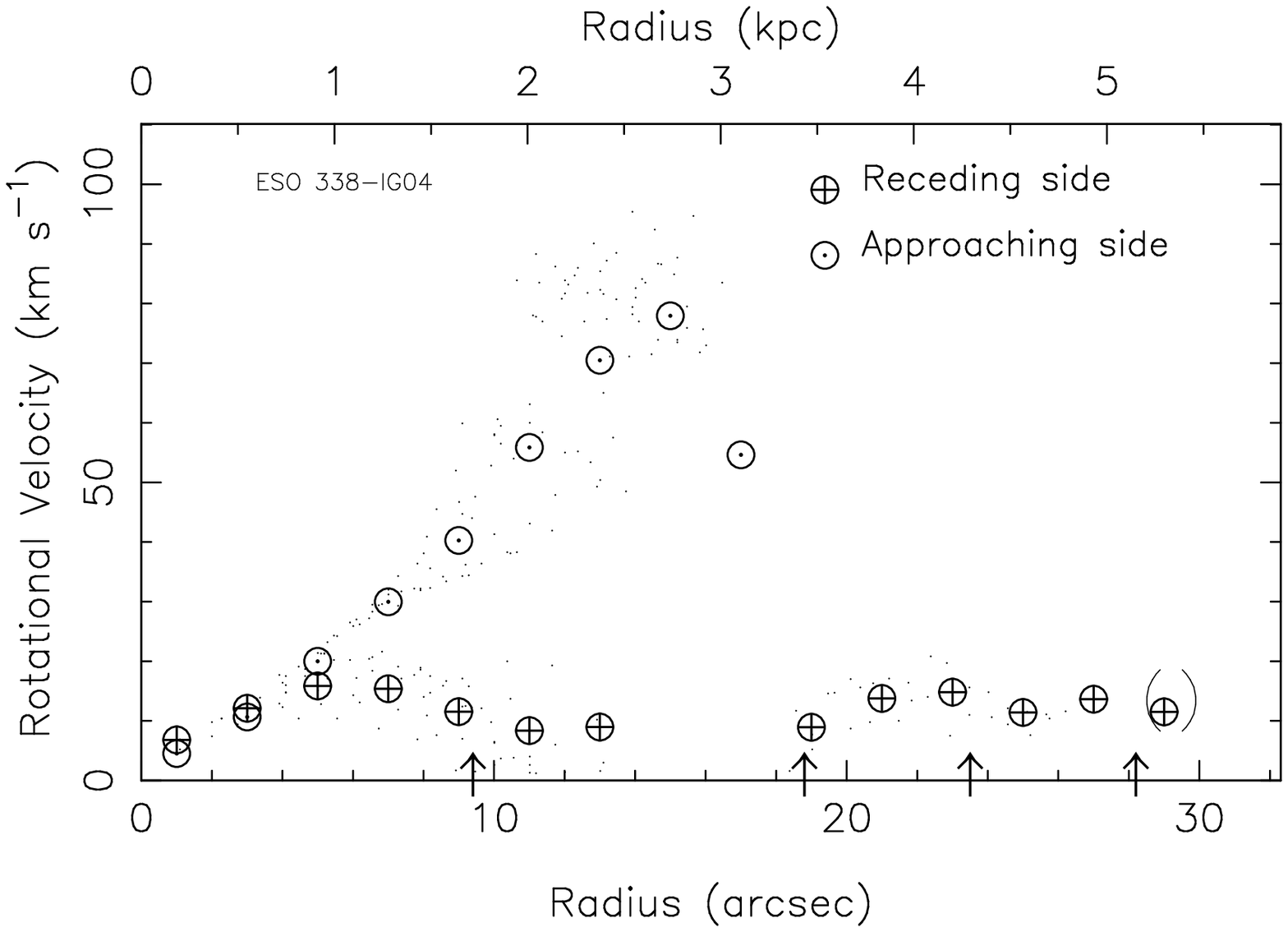}}
\caption{
\object{ESO 338-IG04}:
{\bf Upper left}: 
Continuum isointensity image. Note the bright field star superimposed on the western part of the galaxy.
{\bf Mid-left}: 
Monochromatic (H$\alpha$) isointensity map.  The step between two consecutive levels
is the same for the continuum and monochromatic images.
{\bf Upper right}: 
Isovelocity contours of the ionised gas (thick solid lines)
superimposed on the monochromatic image (thin solid lines).
The step between two consecutive contours is 10 times larger than the
step used in the mid-left image. The spatial scale in each image is shown in the lower right corner. North is up, east is to the left.
{\bf  Bottom}: Rotation curve (RC)  based on $S=  45 \degr$, $incl = 55\degr$~ and ~$PA = 60\degr$.
The arrows 
($\uparrow$) on the abscissa mark the location of $0.4 R_{25}$, $0.8  R_{25}$, 
$1.0  R_{25}$ and $1.2  R_{25}$. Due to the disagreement between the approaching and receding sides, no average RC has been computed. For further explanation of what is shown in the RC see the caption of Fig. \ref{e350} and Sect. 5.
}
\label{e338_1}
\end{figure*}

This galaxy is also well known as \object{Tololo 1924-416} or \object{SCHG 1924-416}. Its photometric properties have been  quite extensively discussed by Bergvall (\cite{bergvall}) and \"Ostlin et al. (\cite{ostlin:e338gc}).  
This galaxy has been observed with two different 
Fabry-Perot interferometers
(FP1 and FP2, see Table 1) and thus we had the opportunity to check the consistency of
the analysis (cf. Fig. \ref{e338_1} and \ref{e338_3}).  First, we successfully 
checked that the flux in the field star superimposed on the western part
of the galaxy and the flux in the H$\alpha$  emission line regions
were consistent between both observations.  Secondly, we checked
that the superimposed bright field star does not seriously affect the 
velocity field. 
Thirdly, we 
confirm that we find the same shape for both velocity fields. 

 The H$\alpha$ emission
is concentrated to an extended bright  central starburst region. In addition there
is diffuse H$\alpha$ emission extending in a tail towards the east. There are also suggestions of a small H$\alpha$ arm emanating towards the south from the western side of the starburst.  
The velocity field is very irregular and does not contain a single axis
of symmetry.  Moreover the velocity gradient is steep
in the western parts and roughly east-west orientated; while in the eastern 
regions the gradient is much lower, not always positive and lacking a well defined position angle. The eastern extended tail has almost no velocity gradient.

East of the centre, just at the border of the starburst region, we observe what appears to be the superposition of two patterns
with different orientation (see Fig. \ref{e338_1}). We will refer to the hypothetical component east of the centre as the {\emph{perpendicular}}   component, since its $PA$ is roughly perpendicular to  the major axis of the galaxy. 
However, we do not observe any 
double component in the profiles (in either FP2 or FP1 data). 
Anyway, we  tried to decompose the velocity field
 without any conclusive results. 
This could mean that if there are two components present, the
linewidth is too large with respect to their velocity separation 
or/and where the components overlap, one only sees the
one with the strongest H$\alpha$ emission.

 In Fig. \ref{e338_3} we show a map of the velocity dispersion as derived from the FWHM linewidth of the FP1 data. The velocity dispersion has a fairly constant level of $\approx 100$ km/s. Where the major axis of the perpendicular component  cross the major axis for the whole galaxy the velocity dispersion is higher and peaks at 160 km/s. Still, the shape of the H$\alpha$ line is consistent with one single broad component.

It is not unambiguous how to derive a RC for this galaxy.  
The different RCs are however consistent in the way that they all
are very irregular, signifying a non equilibrium system. The general feature is an approaching side with continuous steeply rising
velocity, and a receding side with very small velocity gradient. This 
means that the assumption of a regularly rotating disc in equilibrium
must be far from reality in this galaxy. In effect the kinematical centre is not well defined.

\begin{figure*}
\resizebox{16cm}{!}{\includegraphics{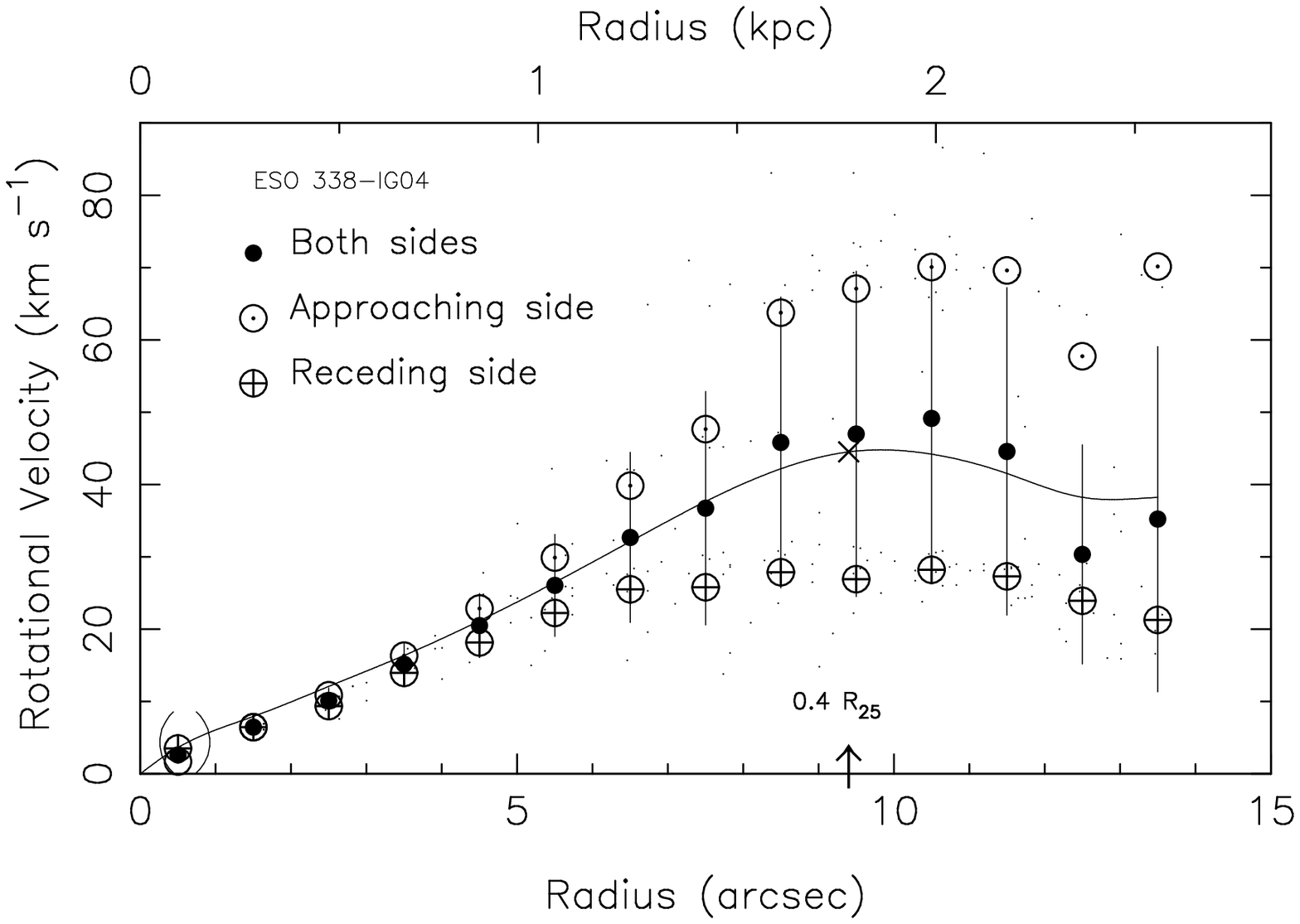}}
\resizebox{16cm}{!}{\includegraphics{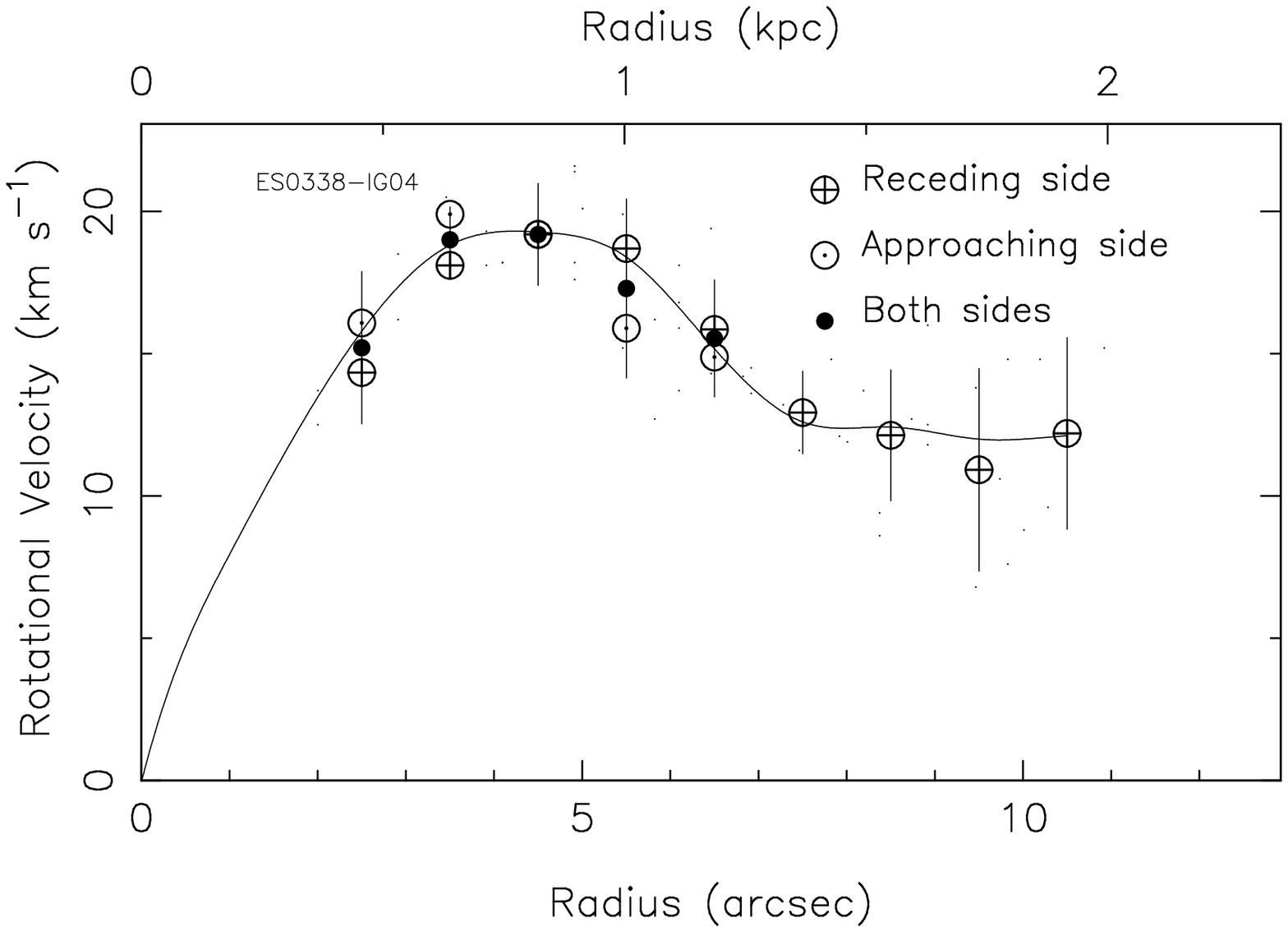}}
\caption{
\object{ESO 338-IG04} continued: {\bf Upper}:
  Rotation curve (RC)  of main component using the ``masked'' velocity field, i.e. with points north-east of the centre and in the eastern tail (up to where the isophotes of the bright central starburst region starts in Fig. \ref{e338_1}) excluded, and limiting the half sector to ~$S = 30\degr$. As before $incl = 55\degr$~ and ~$PA = 60\degr$.  
{\bf Bottom}:  Rotation curve of the "perpendicular" component, based on ~$S=20\degr$, $incl = 62\degr$~ and ~$PA = 336\degr$ . Its centre lies 5$\arcsec$ east of the point where the H$\alpha$ flux has its peak. For further explanation of what is shown in the RCs see the caption of Fig. \ref{e350} and Sect. 5.
}
\label{e338_2}
\end{figure*}

\begin{figure*}
\resizebox{15cm}{!}{\includegraphics{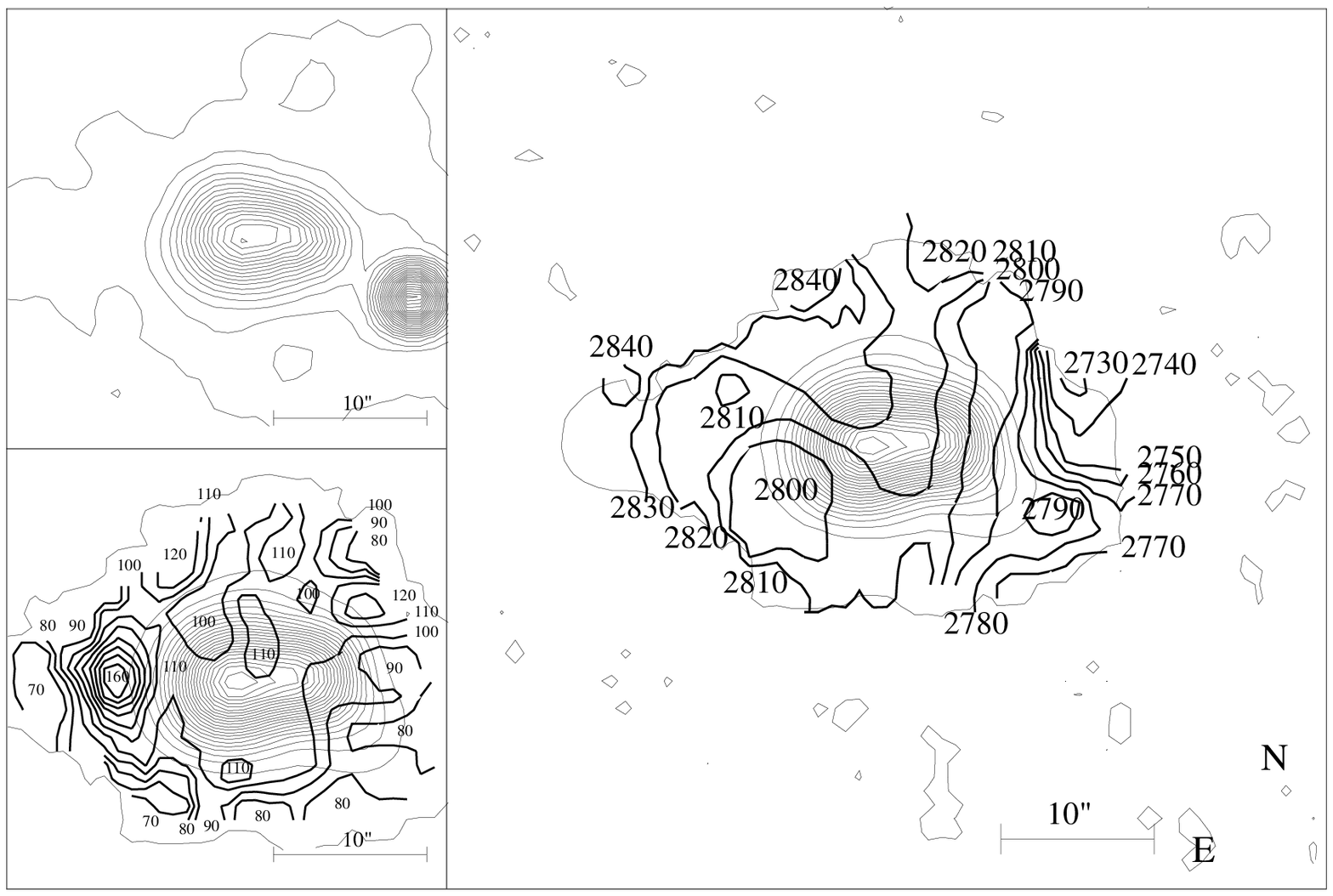}}
\resizebox{\hsize}{!}{\includegraphics{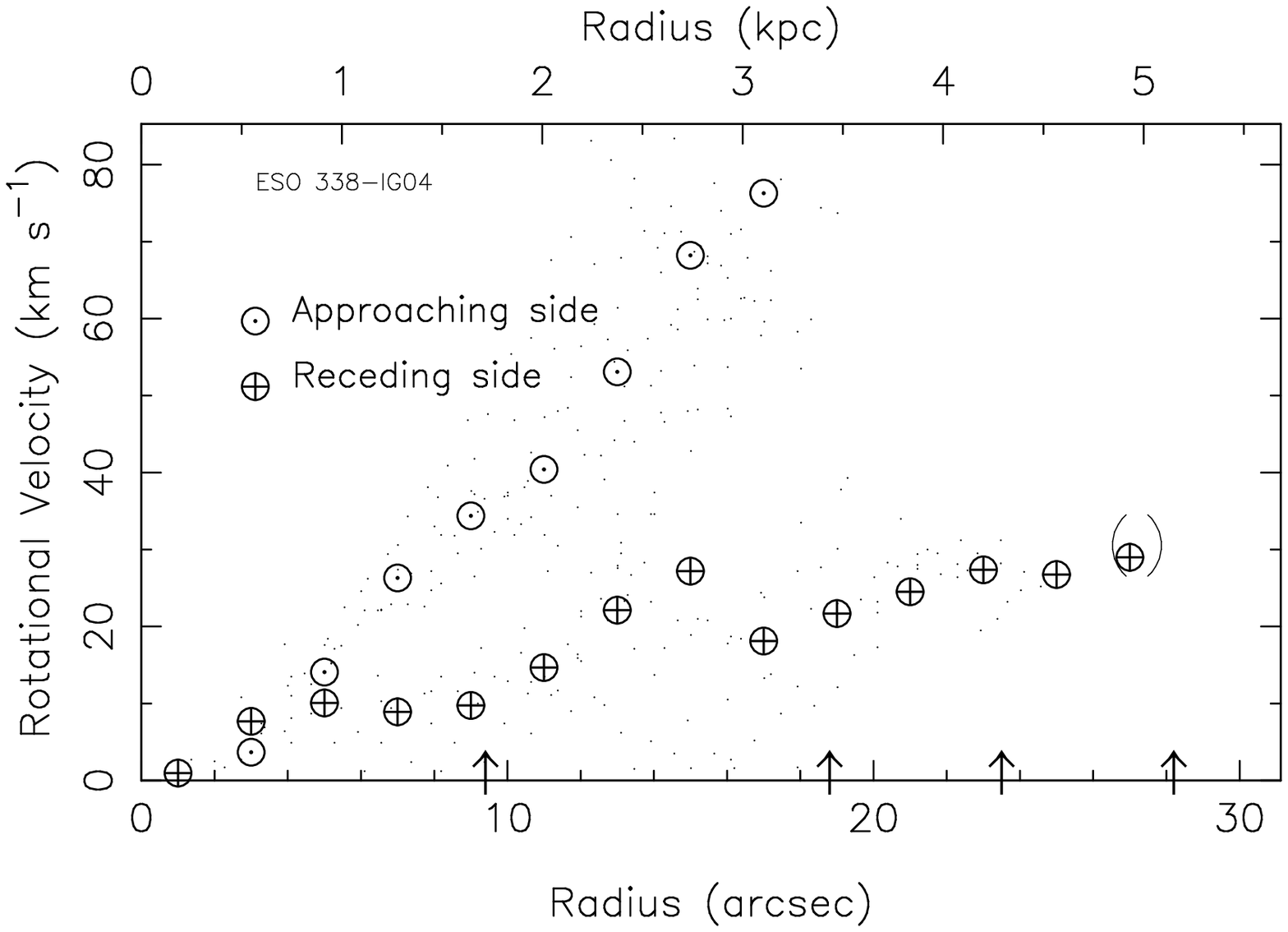}}
\caption{
\object{ESO 338-IG04} continued: Based on FP1 data, with lower velocity resolution:
{\bf Upper left}: 
Continuum isointensity image. Note the bright field star superimposed on the western part of the galaxy.
{\bf Mid-left}: 
Map  of the velocity dispersion (FWHM) in units of km/s.
{\bf Upper right}: 
Isovelocity contours of the ionised gas (thick solid lines)
superimposed on the monochromatic H$\alpha$~ image (thin solid lines).  The spatial scale in each image is shown in the lower right corner. North is up, east is to the left. 
{\bf Bottom}: Rotation curve (RC)  based on the FP1 velocity field and $S = 55\degr$, $incl = 55\degr$~ and ~$PA = 60\degr$. The arrows 
($\uparrow$) on the abscissa mark the location of $0.4  R_{25}$, $0.8 
 R_{25}$, $1.0 R_{25}$ and $1.2  R_{25}$. Due to the disagreement between the approaching and receding sides, no average RC has been computed. For further explanation of what is shown in the RC see the caption of Fig. \ref{e350} and Sect. 5.
}
\label{e338_3}
\end{figure*}

As a cure we tried
to mask away certain points of the velocity field (those in the eastern arm and the approaching side of the perpendicular component) to check if we could obtain a more regular  RC. The result is shown in Fig. \ref{e338_2}a, but  the RC is still far from regular. Moreover, the decline of the average RC outside 2 kpc is faster than the Keplerian case, which is not necessary significant since the velocity field is so perturbed.

For the hypothetical perpendicular
component we could extract a regular RC (Fig. \ref{e338_2}b) only when using a narrow sector
($S < 30 \degr$). The implied rotational velocity is very small, but the regularity
and the agreement between the approaching and receding sides still makes
this component well defined.

 The RC obtained from the FP1 data (Fig. \ref{e338_3}) 
largely agrees with that from FP2 data (Fig. \ref{e338_1}), but the rotational velocity of the receding side
is $\approx 10$ km/s larger for the FP1 data. The differences can be attributed to three effects:  Firstly, the  different spectral resolution of FP1 and FP2.  Secondly, the slightly different choice of 
centre coordinates (an independent determination was made for the FP1 data). Thirdly, the FP1 data is somewhat deeper and reach larger radii, and since the velocity field is deprojected these outer points enter also at smaller radii. In view of this, the agreement is very good.  

A last note on this galaxy is that here it is questionable that the expression "rotation curve" is adequate
since there appears to be no regular rotation present. Still, the presented RCs provide important
information on the complexity of the system. However, the dynamical mass estimate in Table 4 should be viewed sceptically. The ionised gas mass is of the order $10^8 \cal M_{\odot}$.

\subsection{\object{ESO 338-IG04 B}}

\begin{figure*}
\resizebox{15cm}{!}{\includegraphics{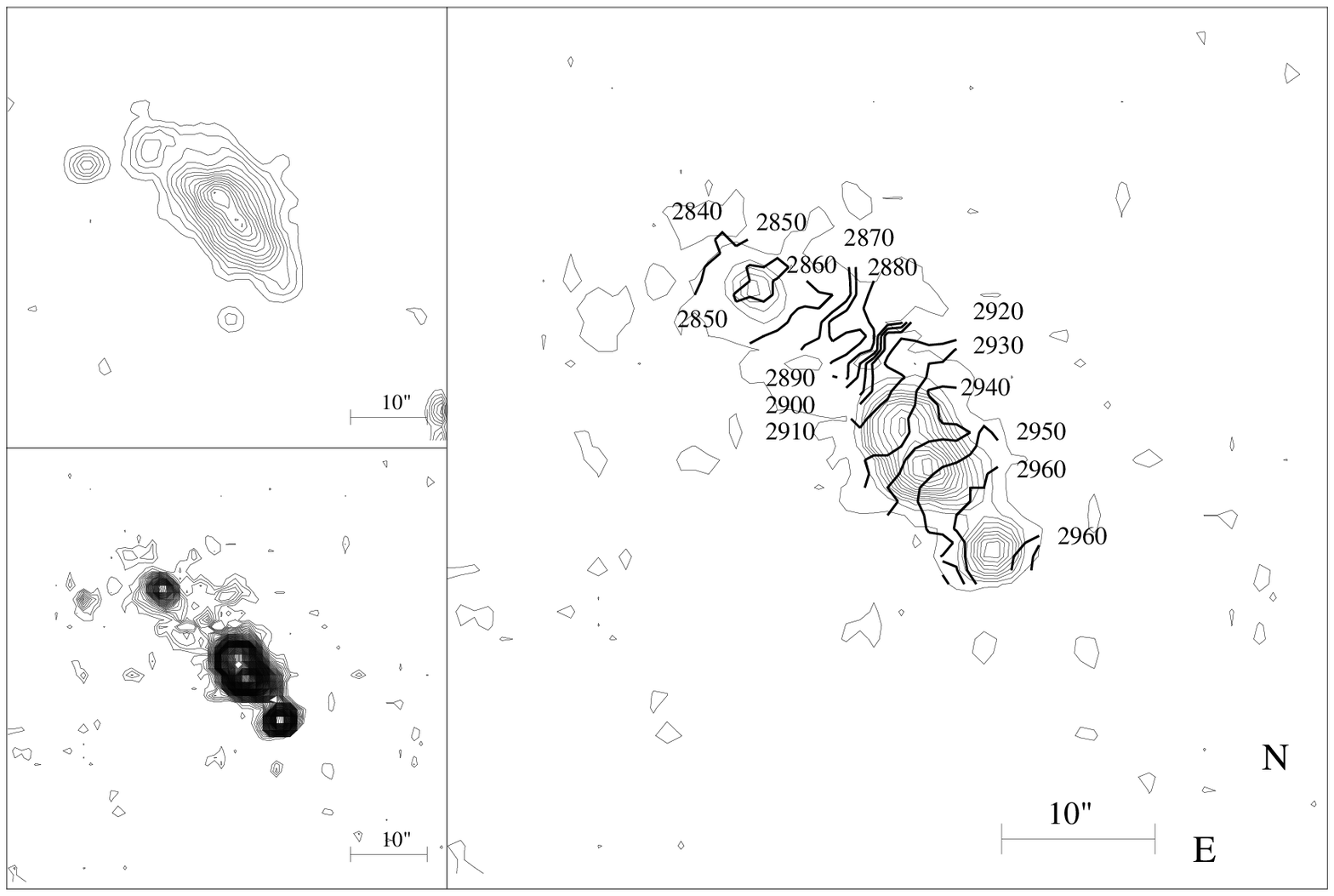}}
\resizebox{\hsize}{!}{\includegraphics{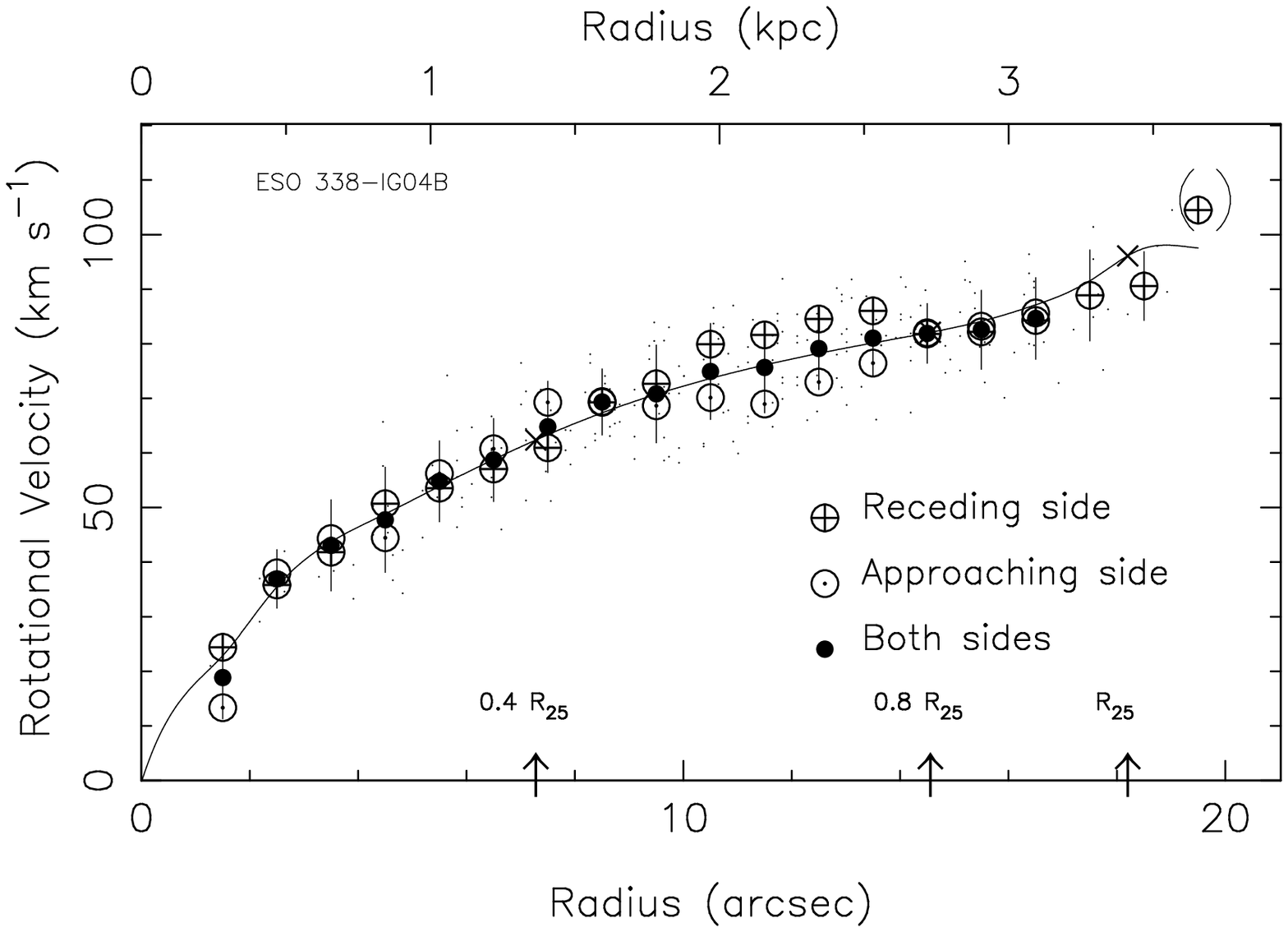}}
\caption{
\object{ESO 338-IG04 B}:
{\bf Upper left}: 
Continuum isointensity image.
{\bf Mid-left}: 
Monochromatic H$\alpha$ isointensity map.  The step between two consecutive levels
is the same for the continuum and monochromatic images.
{\bf Upper right}: 
Isovelocity contours of the ionised gas  (thick solid lines)
superimposed on the monochromatic image (thin solid lines).
The step between two consecutive contours is 10 times lower than the
step used in the mid-left image.  The spatial scale in each image is shown in the lower right corner. North is up, east is to the left. 
{\bf Bottom}: Rotation curve (RC)  based on $S = 40\degr$, $incl = 55\degr$~ and ~$PA = 230\degr$. For further explanation of what is shown in the RC see the caption of Fig. \ref{e350} and Sect. 5.
}
\label{e338b}
\end{figure*}

This galaxy is a physical companion of \object{ESO 338-IG04} and is located approximately six arcminutes south-west of it, corresponding to a projected distance of $\approx 70$ kpc.
The H$\alpha$ image contains at least four bright  \ion{H}{ii} regions (Fig. \ref{e338b}).  
The galaxy has an irregular, somewhat "bent"  shape in H$\alpha$
but the velocity field is still quite regular. Along the north-west side there are twists in the isovelocity contours which may indicate the  presence of a spiral arm. 
Just north-east of the brightest  \ion{H}{ii} region there is a  steep velocity gradient
 which we interpret as the kinematical centre, and which coincides with the centre of broad band images. The RC 
shows good agreement between the two sides and a gradual flattening with increasing radius. 
The estimated dynamical mass is $5 \cdot 10^9 \cal M_{\odot}$ and the ionised gas mass is of the order of $10^7 \cal M_{\odot}$. 

On broad band  CCD images, about  $40 \arcsec$ north-west of the target galaxy, there
is a  small low surface brightness galaxy,  which is not detected in the monochromatic H$\alpha$ images.

\subsection{\object{ESO 185-IG13}}
\begin{figure*}
\resizebox{15cm}{!}{\includegraphics{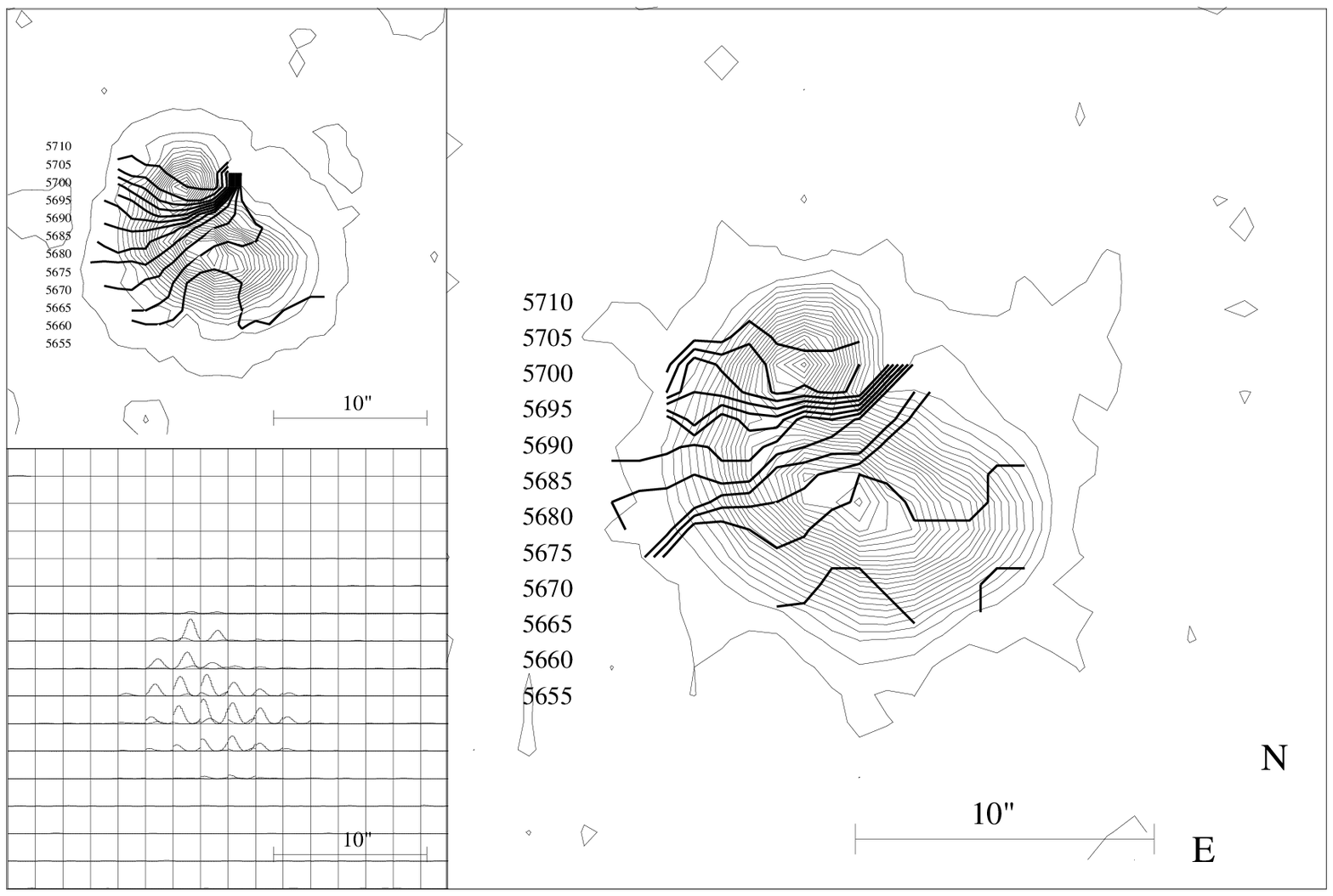}}
\resizebox{\hsize}{!}{\includegraphics{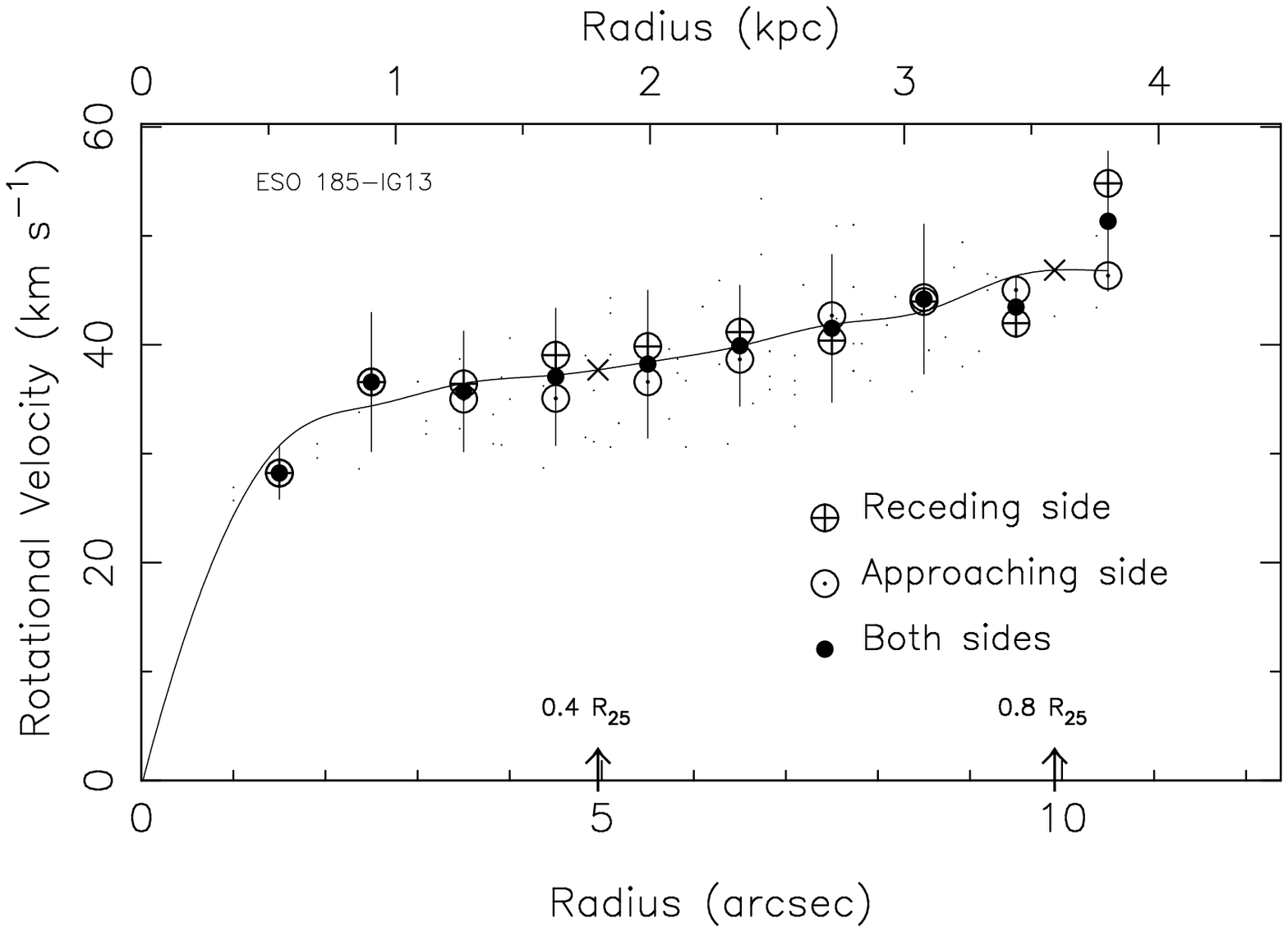}}
\caption{
\object{ESO 185-IG13}:
{\bf Upper left}: 
Total monochromatic H$\alpha$ emission (no decomposition of first and secondary components), with the total velocity field overlaid.
{\bf Mid-left}: 
Profiles within square boxes of ~$2 \times 2$~ pixels of the decomposed first and second monochromatic 
components.
{\bf Upper right}: 
Isovelocity contours of the ionised gas of the first component (thick solid lines)
superimposed on the first monochromatic H$\alpha$ image (thin solid lines).
The step between two consecutive contours is 4 times larger here
than those used for the continuum contours.  The spatial scale in each image is shown in the lower right corner. North is up, east is left.
{\bf  Bottom}: Rotation curve  (RC) of the primary (1st) component based on $S = 40\degr$, $incl = 55\degr$~ and ~$PA = 25\degr$.
For further explanation of what is shown in the RC see the caption of Fig. \ref{e350} and Sect. 5.
}
\label{e185_1}
\end{figure*}

This galaxy has a nearly round shape in the continuum but a more complex H$\alpha$ morphology and  velocity field (Fig. \ref{e185_1}). Broad band images reveal arms and a $\ge 30 \arcsec$ long tail extending to the north-east.
The H$\alpha$ emission is strong over a large area with two peaks  in  the surface brightness. The velocity field appears squeezed in the north-west region, 
between the two peaks in the H$\alpha$ surface brightness. About 1.5 arcminutes south-west of the galaxy there are two galaxies present, none
of which is detected in H$\alpha$ at the redshift of \object{ESO 185-IG13}. 

\begin{figure*}
\resizebox{15cm}{!}{\includegraphics{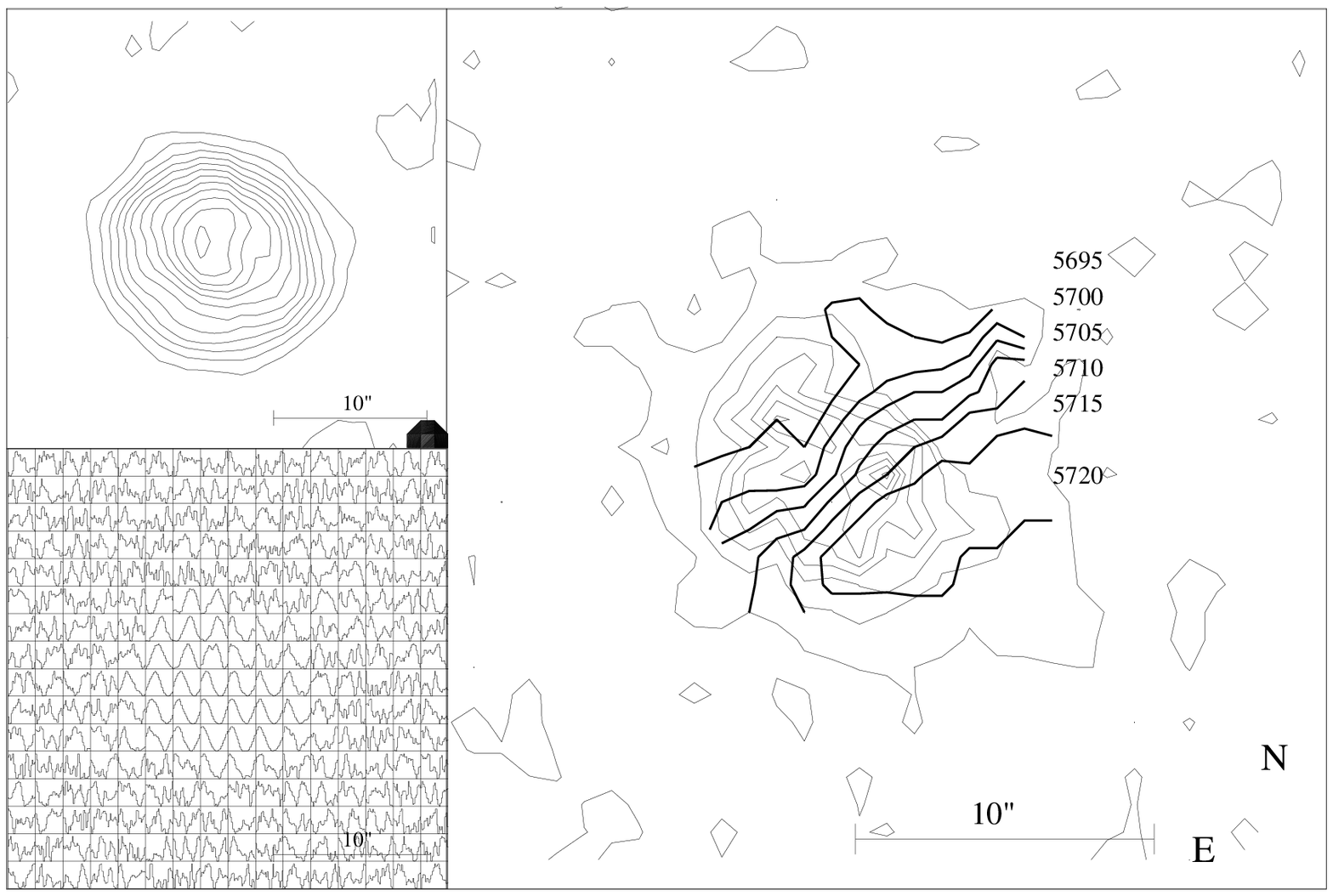}}
\resizebox{\hsize}{!}{\includegraphics{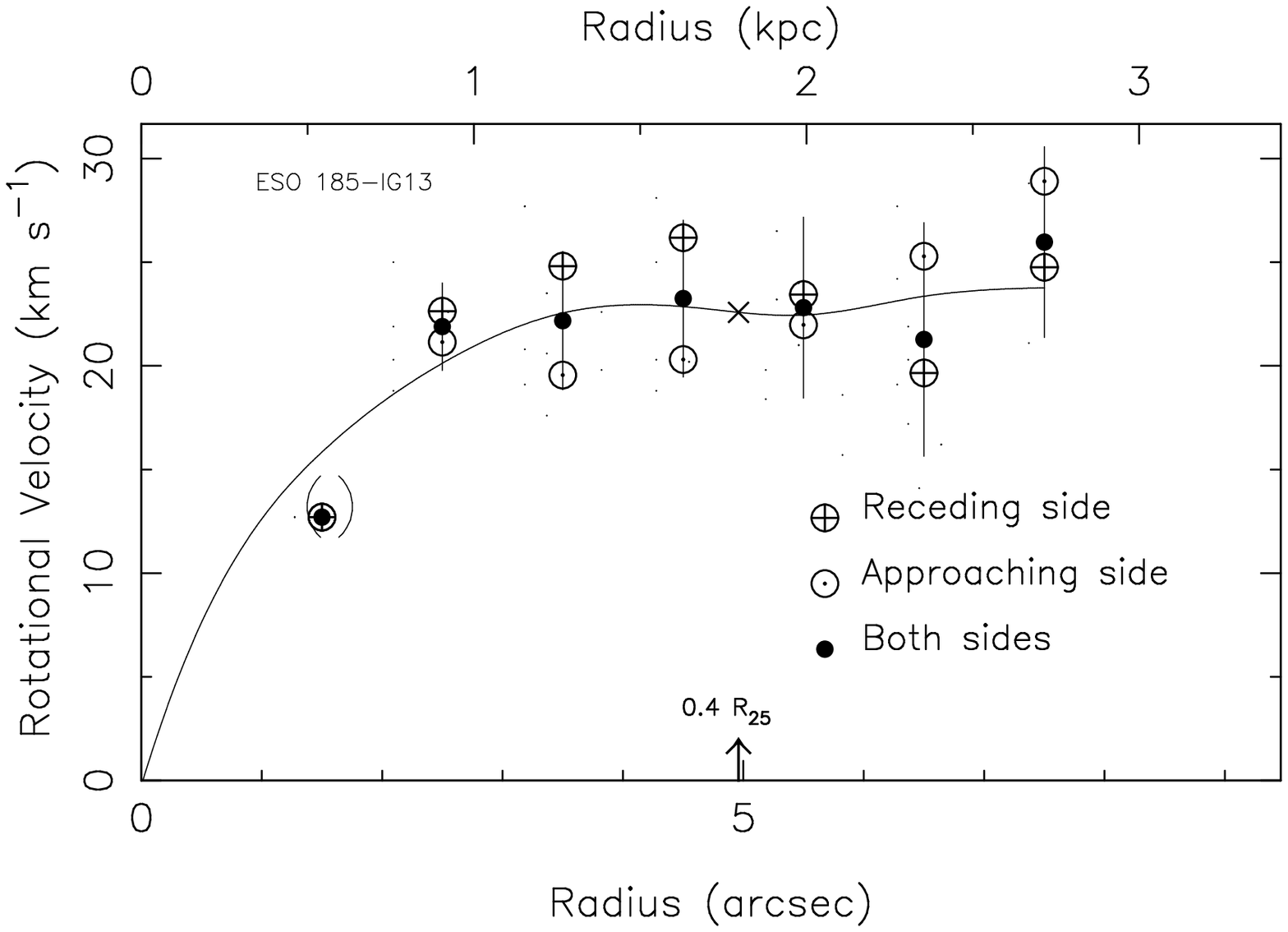}}
\caption{ 
\object{ESO 185-IG13} continued:
{\bf Upper left}: 
Total continuum isointensity image.
{\bf Mid-left}:
Line profiles  
of the non-decomposed data plotted within squares of ~$2 \times 2$~ pixels. The intensity scale (Y-axis) in each 
box with line profiles is normalised to the peak intensity in that box.
{\bf Upper right}: 
Isovelocity contours of the ionised gas of the second component (thick solid lines)
superimposed on the second monochromatic H$\alpha$ image (thin solid lines).
The step between two consecutive contours is 4 times larger here
than those used for the continuum contours but is the same as the
step chosen for the first component.  The spatial scale in each image is shown in the lower right corner. North is up, east is to the left. 
{\bf Bottom}: Rotation curve (RC)  of the secondary component based on $S = 40\degr$, $incl = 50\degr$~ and ~$PA = 225\degr$.
For further explanation of what is shown in the RC see the caption of Fig. \ref{e350} and Sect. 5.
}
\label{e185_2}
\end{figure*}

Double components with rather different intensity have been extracted
from the original  lines.  Both components have
similar position angles but the gradients are reversed, i.e. the secondary component is counter-rotating with respect to the primary (Fig. \ref{e185_1} and \ref{e185_2}).
The brightest component is very similar to the total non-decom\-posed
velocity field (compare the upper left and right panels in Fig. \ref{e185_1}). 
The second  component is more diffuse and regular than the first
one.  
 While the centre of the secondary component coincides with the continuum peak, the kinematical centre of the primary
component is slightly offset from it ($\approx 1 \arcsec$ to the north).

The RC of the primary component shows a rapid increase
followed by an almost flat slowly rising part. The agreement between 
the approaching and receding sides is good. The secondary component
also yields a regular RC, with a low rotational velocity
though. The regularity of the RC of the secondary component suggests that it 
is caused by a dynamically well defined object, e.g. a counter rotating disc.
The estimated dynamical mass of \object{ESO 185-IG13} is $\sim 2 \cdot  10^9 \cal M_{\odot}$ and the ionised gas mass is of the order of $10^8 \cal M_{\odot}$. The dynamical mass of the secondary component is $\sim 3 \cdot  10^8 \cal M_{\odot}$

\subsection{\object{ESO 400G-43}}
This galaxy presents several bright \ion{H}{ii}  regions. 
Note that the northernmost region is not a field star. The southern  \ion{H}{ii} region complex has a ring-like morphology.
 In the southern  \ion{H}{ii} region complex we observe a symmetric and regular velocity field with velocity plateau's followed by  decreasing velocity on both sides, characteristic of a disc with circular differential rotation. 
At larger radius ($\ge 5 \arcsec$) the north-east and south-west regions start to behave differently. In the south-west we observe the 
continuity of the central velocity field, although the isovelocity contours are somewhat boxy.
The region north-east of the centre is more intriguing because we observe a shift of the major axis
position angle, rotating from approximately ~45$\degr$~ towards north, which might indicate
the presence of a warp. An alternative interpretation is that the twisted isovelocity contours north of the north-east velocity plateau is due to the presence of a spiral arm. The isovelocity contours in the northern part of the velocity field
seems to indicate the presence of local motions, and asymmetric line profiles are present. The kinematical centre is
well defined  but is slightly offset ($\approx 1 \arcsec$ ~ towards south-west) from the peak continuum emission. 
Due to the irregular velocity field in the north east, the inclination is quite uncertain for this galaxy.  
 
The RC has a strange behaviour. Following the
initial solid body rise to the plateau's, there is a rapid decline after which the RC levels out and stays flat out to a radius of 15 arcseconds, after which the approaching side (the only one that has H$\alpha$ signal) declines further. The remarkable thing is that after the maximum rotational velocity the decline is faster than for the Keplerian case. This is unphysical for an equilibrium disc, thus indicating that the velocity field is distorted. The  ``super-Keplerian'' decline can be eased somewhat, but not completely, by allowing the position angle to be variable. 
In the south west regions at a radius greater than $5\arcsec$ double features in the H$\alpha$ lines appear at low S/N. However, this feature is present in many consecutive pixels and is in total significant, which means that a secondary component might be present here. In the centre there are no signs of double lines and our attempts to decompose the velocity field failed.  Due to the super-Keplerian behaviour of the RC the  estimated mass apparently decreases with increasing radius (see Table 4). Thus the mass stated in Table 4 is very uncertain.
The ionised gas mass is of the order of a few times $10^8 \cal M_{\odot}$, comparable to the dynamic estimate.

\subsection{\object{ESO 400G-43 B}} 
This quite low surface brightness galaxy (Fig. \ref{e400b}) is a physical companion of \object{ESO 400G-43} and is located  approximately three arcminutes east/south-east of it (Berg\-vall and J\"ors\"ater \cite{ber:jo}), corresponding to a projected distance of $\approx 70$kpc.
ESO\-~400G-43B has one extended central emission line region, with peak intensity in the north-west.
  The isovelocity contours shows axisymmetric twists south-east and (to a lesser extent) north-east of the major axis, perhaps indicating the presence of spiral arms. 
The shape of the RC  may be affected by dust extinction, due to the high inclination. However, the best fitting kinematical inclination is significantly lower than the photometric inclination derived from ellipse-fitting of the outer isophotes, the reason for which is not well understood. In the presented RC we have applied a small correction for internal extinction in the central parts, see Fig. \ref{e400b}. The estimated dynamical mass is $\sim 3 \cdot 10^9\cal M_{\odot}$ and the ionised gas mass is of the order of a few times $10^7 \cal M_{\odot}$.

\subsection{\object{Tololo 0341-407}}
Even if this object has a short observing time (see Table 2) it was  possible to extract a
high S/N velocity field. 
Our data indicate that this object is in fact 
composed of two well separated galaxies: the eastern one being more luminous in  H$\alpha$ than the western one (Fig. \ref{tol_1} and \ref{tol_2}). The velocity field of the eastern component displays a  rather constant velocity indicating a close to face on orientation. 
The western component is more inclined and shows more obvious rotation.
Asymmetries in the line profiles suggested a decomposition of the velocity field 
into two components,  one which is dominated by the eastern galaxy and the other by the western one (Fig. \ref{tol_1} and \ref{tol_2}). If the connection between the first component in the eastern and western galaxy is physical is not certain; and the same is  true for the secondary components. Here, the notation with a primary and secondary component may be somewhat confusing, since the secondary component of the western galaxy is actually the most luminous. Hereafter, when we mention the ``first'' component of the eastern or western galaxy we refer to the strongest (most luminous) component, and vice versa for the secondary components.
The secondary component of the eastern galaxy has a constant 
velocity, which suggests that this component  belongs to
the  same galaxy, although the  velocity is more similar the western galaxy.   
The second component
of the western galaxy has a major axis position angle roughly
perpendicular to the one of the first component, which suggest that
it is dynamically distinct from it.
The primary component of the eastern galaxy  has a solid body RC with quite high dispersion. The primary component of the western galaxy  gives a  well determined RC, perhaps levelling out at large
radii. The kinematical centres roughly coincide  with
the peaks in the continuum emission, although this is hard to determine for the
eastern galaxy. The secondary components do not produce regular RCs. 
The eastern and western galaxies both have estimated dynamical masses around $\sim 2 \cdot10^8 \cal M_{\odot}$, and the ionised gas masses are of the order of a few times $10^7  \cal   M_{\odot}$.

\section{Summary}
We have presented velocity fields for a sample of six relatively luminous blue 
compact galaxies, and two companion galaxies. In general the velocity fields, in spite of high S/N,  appear irregular
and distorted, except for the two companion galaxies included in 
the sample. We have also shown the derived rotation curves (RCs) for the
simplest choice of input parameters. That some RCs appear strange
or non-uniform indicates that the simple assumption of a regular
warp-free disc with circular rotation  is not valid in general. Still the information in the
RCs are valuable to get a feeling of the dynamics of these systems
and to get a rough  estimate of the mass contained within the observed extent
of the galaxies. The estimated masses range 
from a few times $10^8$ to $\sim 10^{10} {\cal M_{\odot}}$. 
Secondary  components have smaller estimated masses (when a RC could be derived). These mass estimates are approximate and based on the assumption that the galaxies are supported by rotation alone. We will show in Paper II that some of the galaxies cannot be primarily rotationally supported. 

Using  the  integrated H$\alpha$ fluxes, the masses of ionised gas were  estimated to lie in
the range 10$^7$ to 10$^9 {\cal M_{\odot}} $. This amounts to between less than 1\% and more than 50\% of the estimated dynamical masses. The two companion galaxies had the lowest fraction of ionised gas mass to dynamical mass, and the galaxies with the most peculiar RCs (\object{ESO 350-IG38}, \object{ESO 338-IG04} and \object{ESO 400-G43}) apparently have the largest fraction of their mass in the form of ionised gas.   

This paper presents the most extensive study of the central dynamics and optical
velocity fields of BCGs as yet. 
In addition, all the observations have been done with the same instrument and the
sample selection, reductions and analysis are homogeneous.
In a forthcoming paper (Paper II)
we will discuss the interpretation of the velocity fields and rotation curves in terms 
of  more detailed mass models. In addition we will discuss what the results presented 
here tell us about the triggering mechanism for the starburst activity
present in these galaxies.

\medskip
\begin{acknowledgements}

This work was partly supported by the Swedish Natural 
Science Research Council. 
We thank 
Jean-Luc Gach from Marseille Observatory 
for mounting and dismounting 
the instrument on the telescope and for assistance during the observations.
Eva \"Orndahl is thanked for useful comments on the manuscript. Anna Westman is thanked for her help in editing the text. We would also like to thank Albert Bosma for stimulating discussions on the interpretation of velocity fields.
This research has made use of the NASA/IPAC Extragalactic Database (NED)which is operate by the Jet Propulsion Laboratory, California Institute of Technology, under contract with the National Aeronautics and Space Administration.

\end{acknowledgements}

%
%

%
%

%
%

%
%

%
%

%
%

%
%

%
%
 
%
%

%
%

%
%
\begin{figure*}
\resizebox{15cm}{!}{\includegraphics{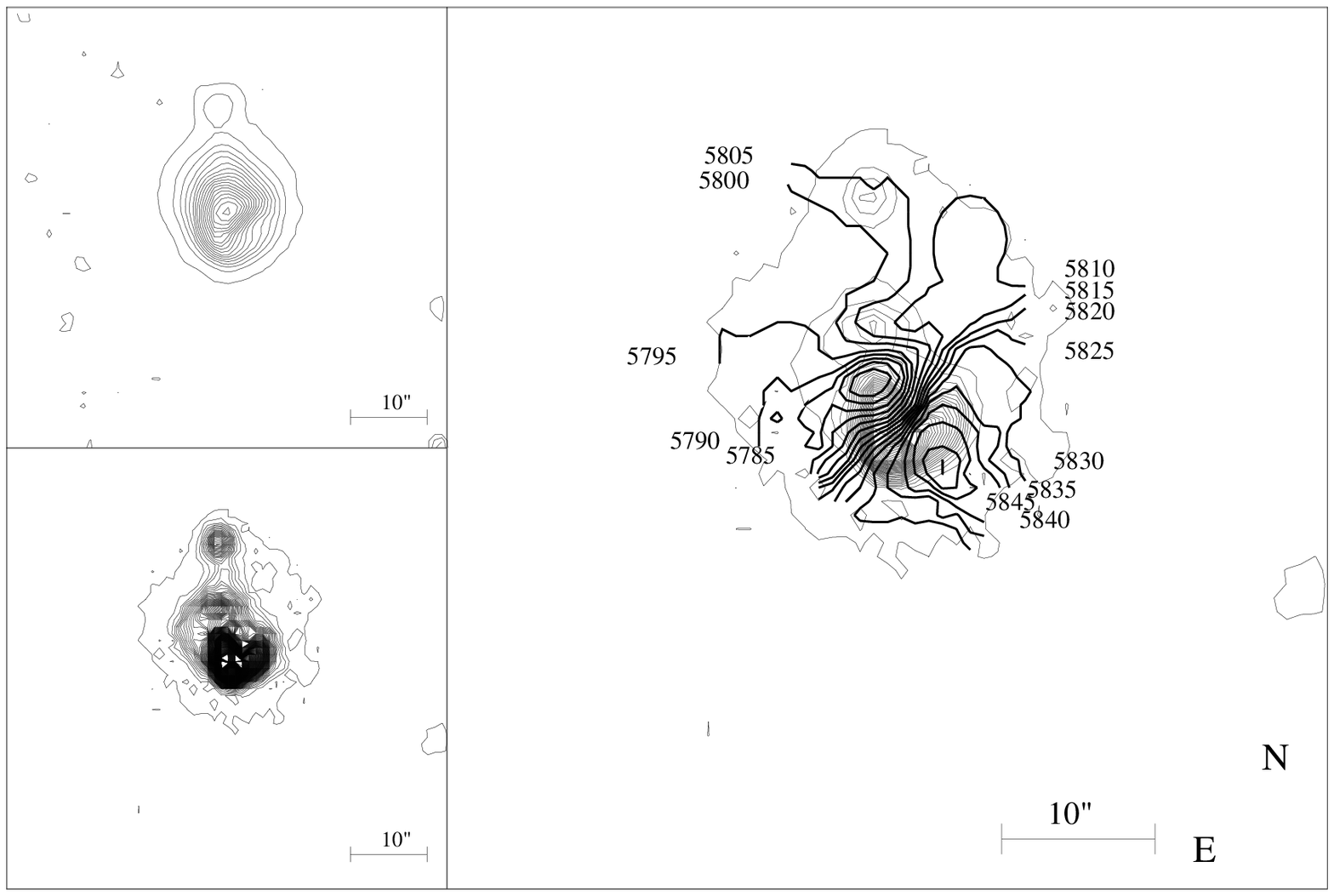}}
\resizebox{\hsize}{!}{\includegraphics{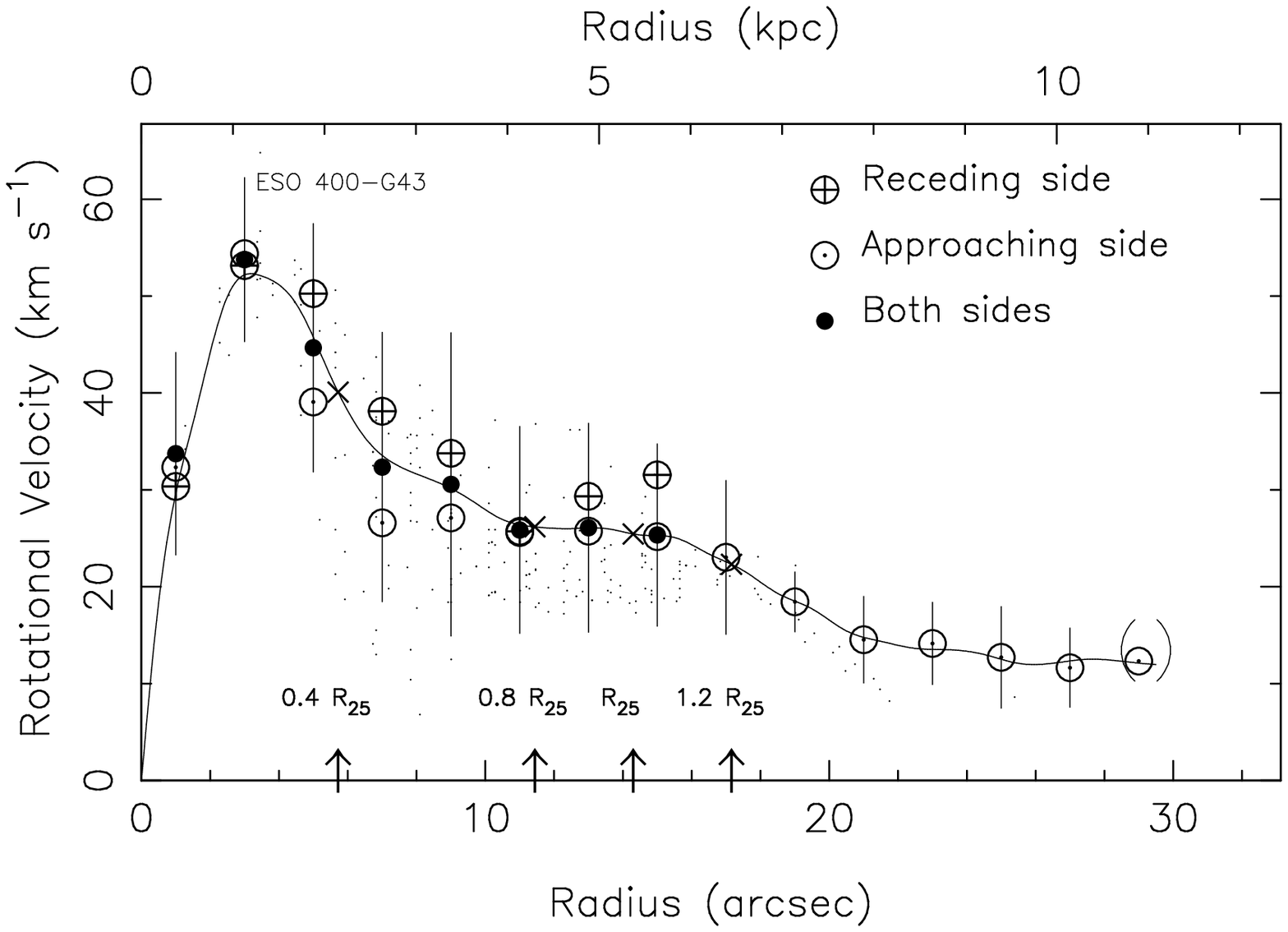}}
\caption{
\object{ESO 400-G43}:
{\bf Upper left}: 
Continuum isointensity image. 
{\bf Mid-left}: 
Monochromatic H$\alpha$ isointensity map.  The step between two consecutive levels
is the same for the continuum and monochromatic images.
{\bf Upper right}: 
Isovelocity contours of the ionised gas (thick solid lines)
superimposed on the monochromatic image (thin solid lines).
The step between two consecutive contours is 6.7 times larger than the
step used in mid-left image.
{\bf  Bottom}: Rotation curve (RC)  based on $S = 55\degr$, $incl = 55\degr$~ and ~$PA = 225\degr$.
For further explanation of what is shown in the RC see the caption of Fig. \ref{e350} and Sect. 5.
}
\label{e400}
\end{figure*}

%
%
\begin{figure*}
\resizebox{15cm}{!}{\includegraphics{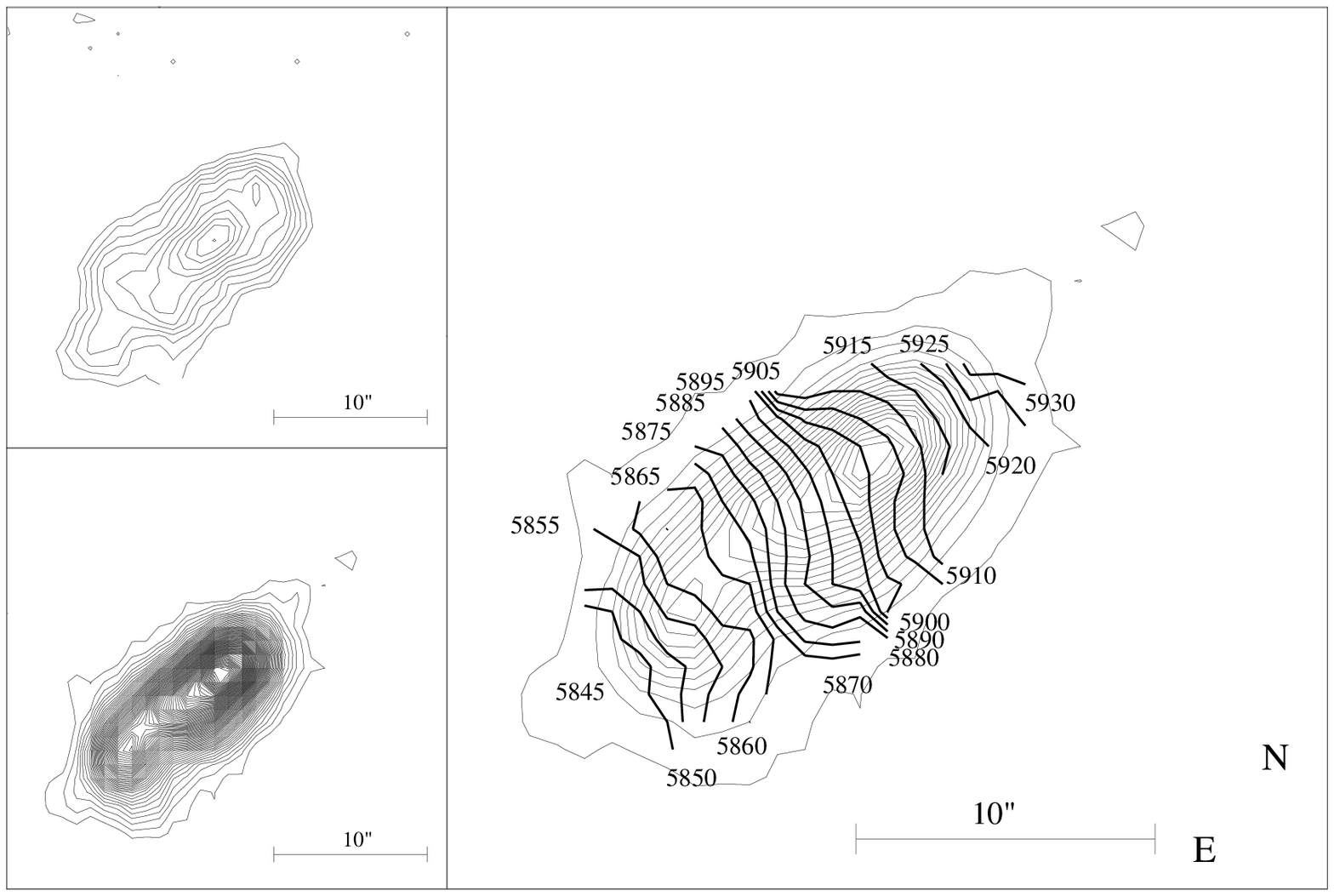}}
\resizebox{\hsize}{!}{\includegraphics{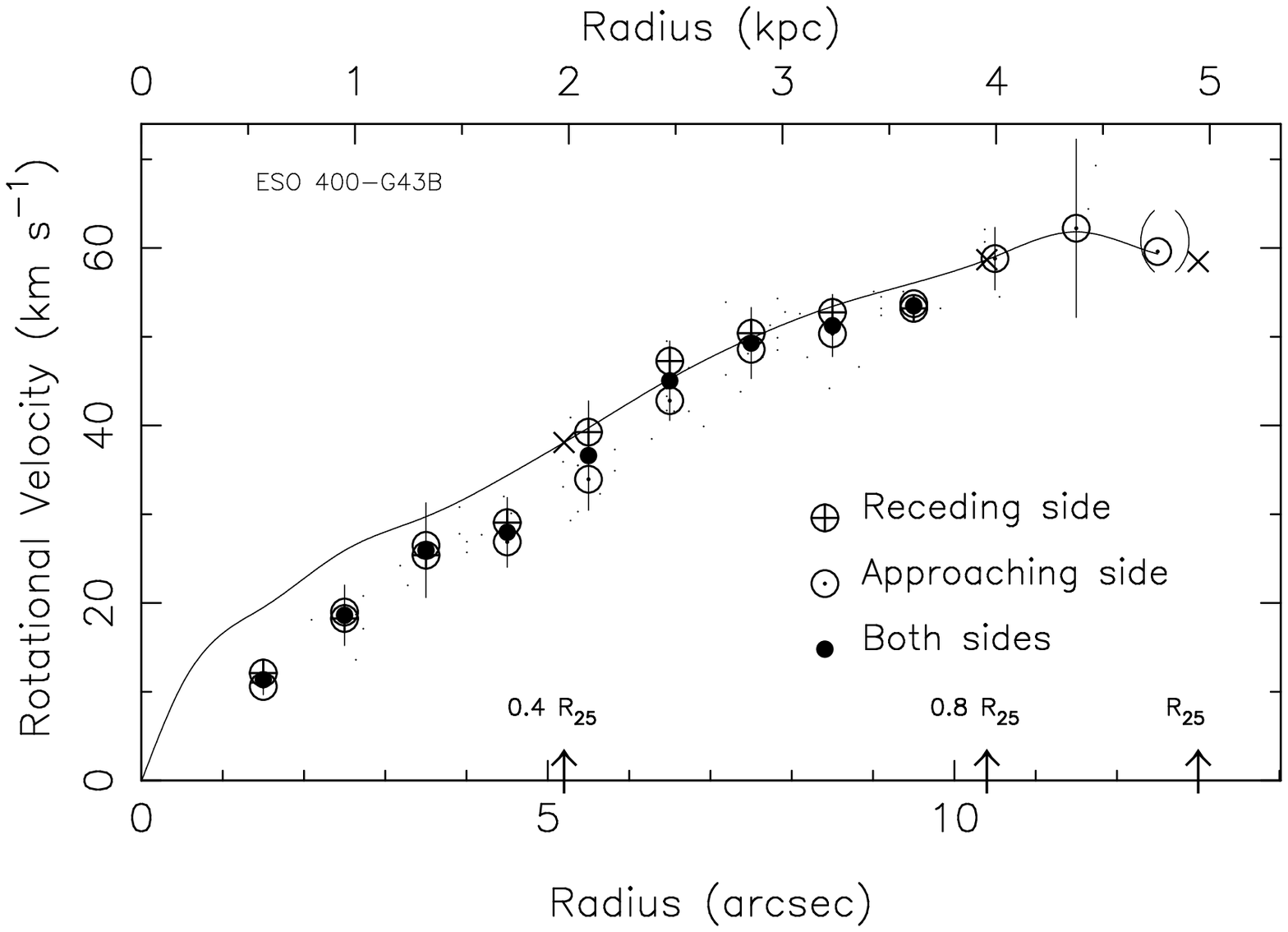}}
\caption{
\object{ESO 400-G43 B}:
{\bf Upper left}: 
Continuum isointensity image.
{\bf Mid-left}: 
Monochromatic H$\alpha$ isointensity map.  The step between two consecutive levels
is 5 times larger for the monochromatic image than for the continuum map.
{\bf Upper right}: 
Isovelocity contours of the ionised gas (thick solid lines)
superimposed on the monochromatic image (thin solid lines).
The step between two consecutive contours is 4 times larger than the
step used in mid-left image.  The spatial scale in each image is shown in the lower right corner. North is up, east is to the left. 
{\bf Bottom}: Rotation curve (RC)  based on $S = 45\degr$, $incl = 60\degr$~ and ~$PA = 305\degr$. The fitted curve (solid line) has been corrected for absorption in the centre due to the high inclination of this galaxy, therefore the curve lies slightly above the measured velocity points at radii smaller than $\arcsec$ 
For further explanation of what is shown in the RC see the caption of Fig. \ref{e350} and Sect. 5.
}
\label{e400b}
\end{figure*}

%
%
\begin{figure*}
\resizebox{15cm}{!}{\includegraphics{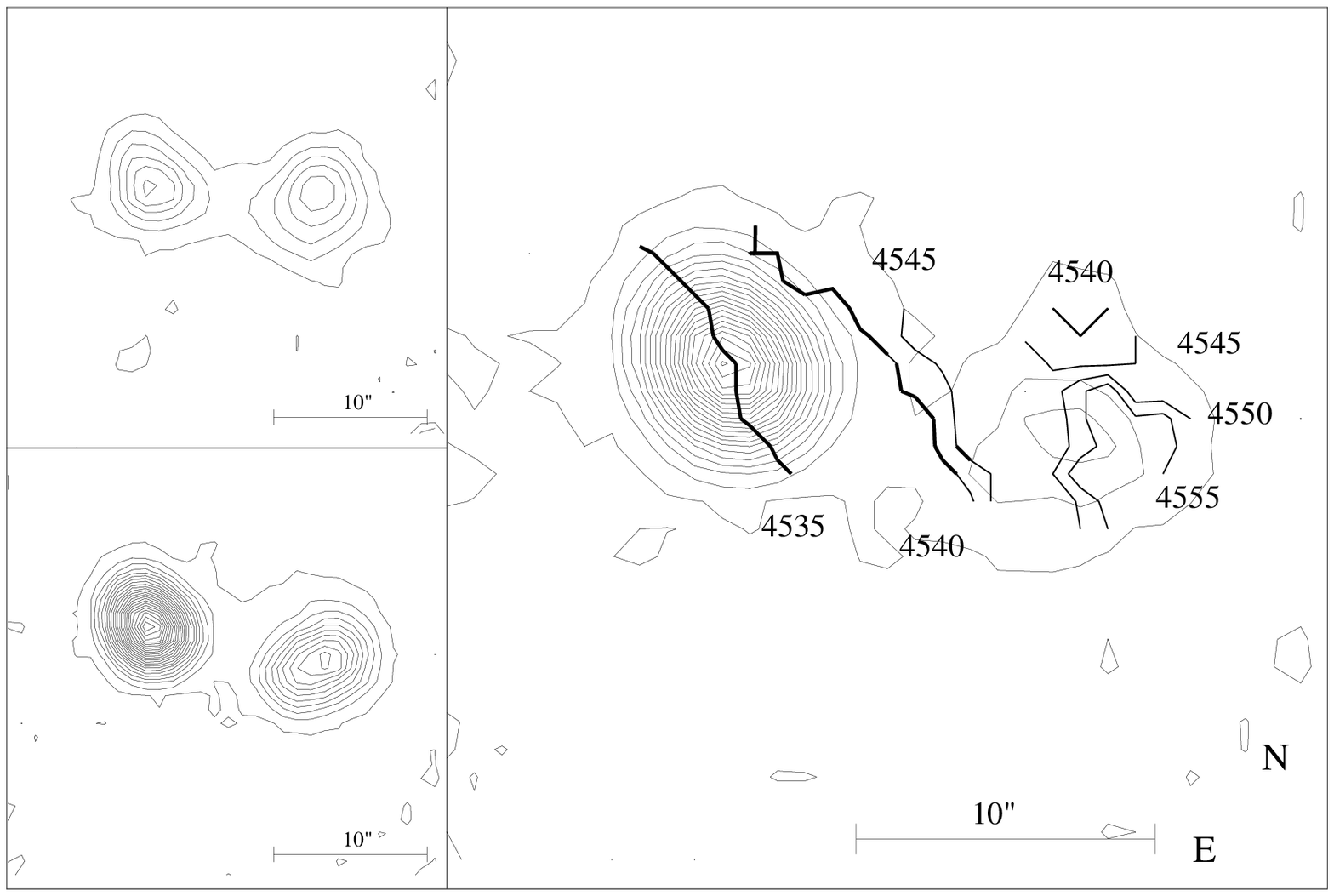}}
\resizebox{\hsize}{!}{\includegraphics{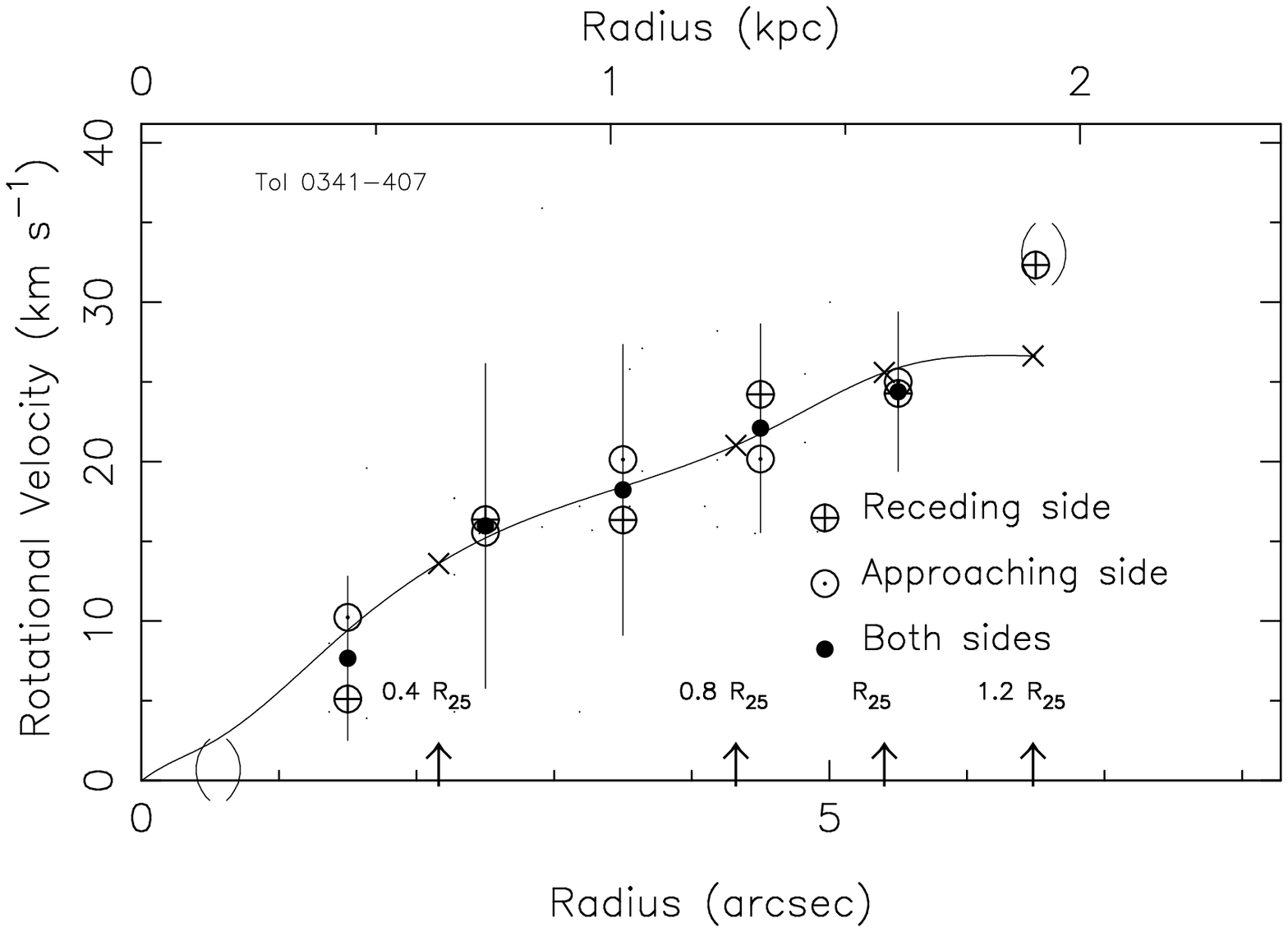}}
\caption{
\object{Tololo 0341-407}:
{\bf Upper left}: 
Continuum isointensity image.
{\bf Mid-left}: 
Total H$\alpha$ emission (addition of both components).
{\bf Upper right}: 
Isovelocity contours of the ionised gas (thick solid lines) of the first component of the eastern galaxy, and the isovelocity contours of the second component (thin lines) of the western galaxy
superimposed on the monochromatic H$\alpha$ image of the first component (where the eastern galaxy dominates).
The step between two consecutive contours is the same 
as those used for the continuum contours. The spatial scale in each image is shown in the lower right corner. North is up, east is to the left.
 {\bf Bottom}: Rotation curve (RC)  of the eastern galaxy first component, based on ~$S = 60 \degr$, $incl = 15\degr$~ and ~$PA = 295\degr$.
For further explanation of what is shown in the RC see the caption of Fig. \ref{e350} and Sect. 5.
}
\label{tol_1}
\end{figure*}

%
%

\begin{figure*}
\resizebox{15cm}{!}{\includegraphics{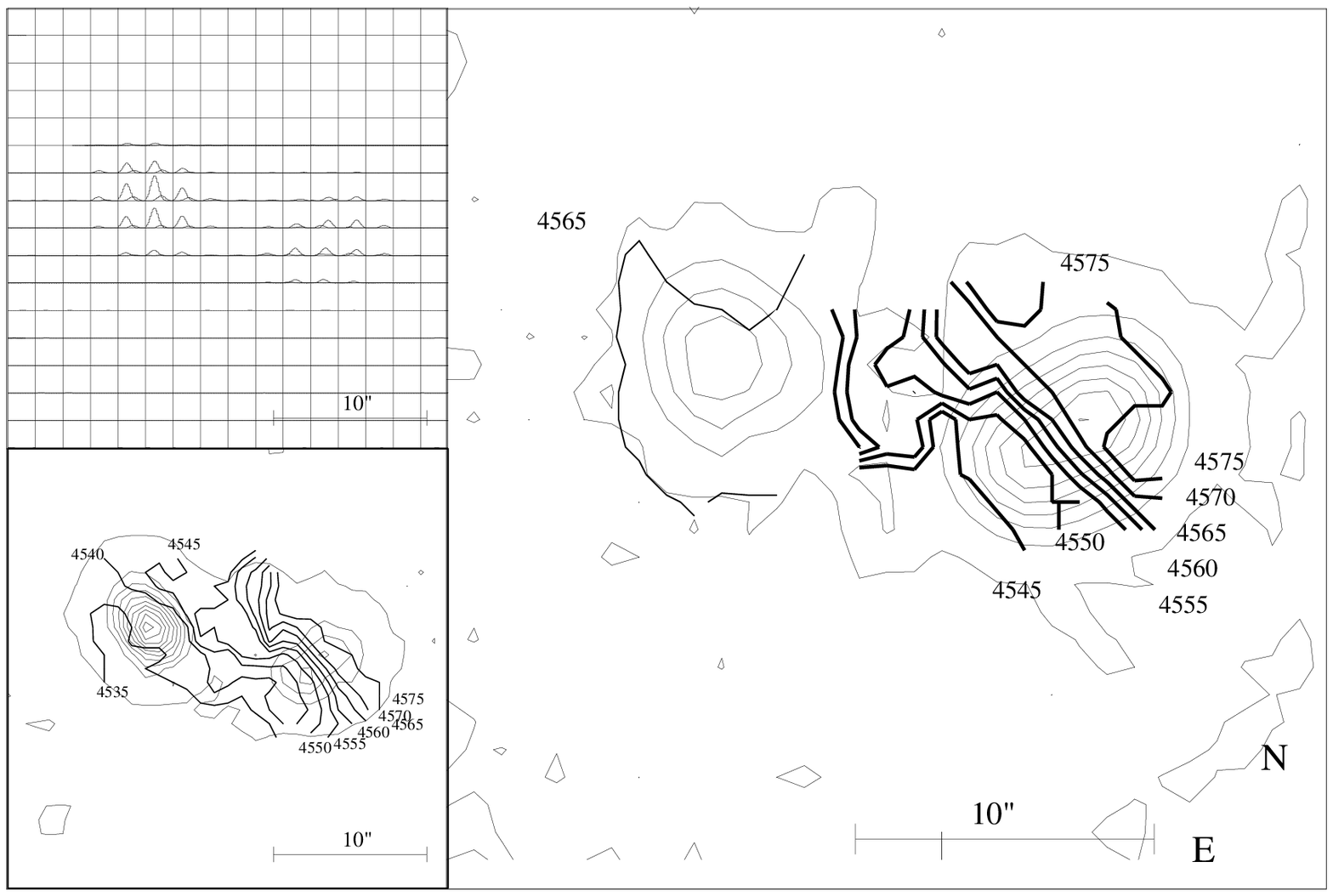}}
\resizebox{\hsize}{!}{\includegraphics{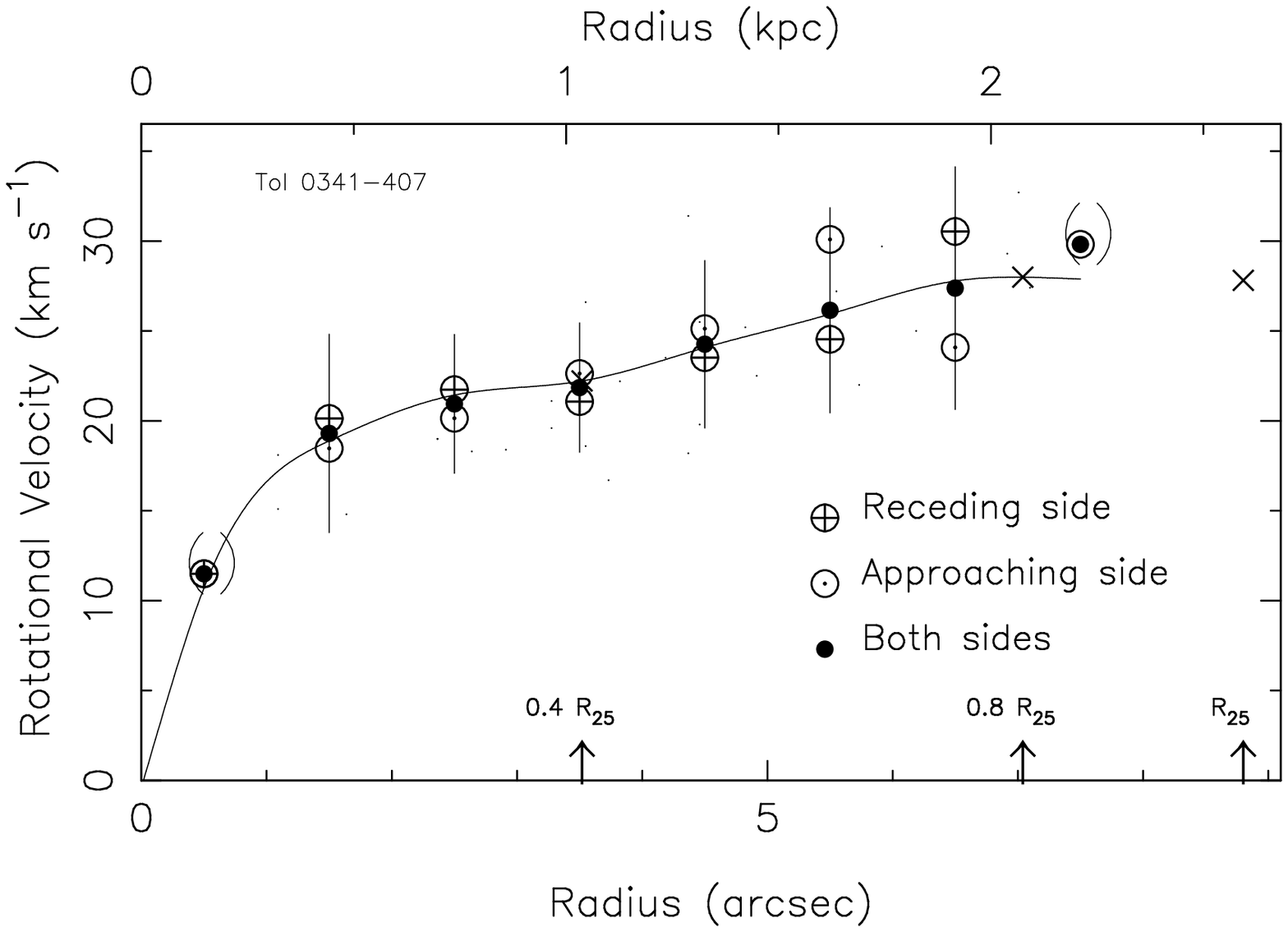}}
\caption{
\object{Tololo 0341-407} continued:
{\bf Upper left}: 
Individual profiles.
Two Gaussian profiles are plotted in each pixel,  
corresponding to the two monochromatic components and the two velocity
fields.  
{\bf Mid-left}: 
Total H$\alpha$ emission (addition of both components) 
with total (non-decomposed) velocity field overlaid. 
{\bf Upper right}: 
Isovelocity contours of the ionised gas of the first component of the western galaxy (thick solid lines) and isovelocity contours of the second component of the eastern galaxy (thin lines);
superimposed on the monochromatic H$\alpha$ image of the second component (where the western galaxy dominates).
The step between two consecutive contours is the same 
as that used for the continuum and first monochromatic contours. The spatial scale in each image is shown in the lower right corner. North is up, east is to the left.  
{\bf Bottom}: Rotation curve (RC) of the first component of the western galaxy, based on ~$S = 50 \degr$, $incl = 50\degr$~ and ~$PA = 313\degr$. For further explanation of what is shown in the RC see the caption of Fig. \ref{e350} and Sect. 5. 
}
\label{tol_2}
\end{figure*}

\end{document}